\documentclass[a4paper,11pt]{article}
\usepackage{jheppub} 
\usepackage{lineno,braket,booktabs,makecell,nicematrix}
\usepackage{braket,soul}
\usepackage{ulem}
\usepackage{amsmath,cleveref}
\usepackage{amsfonts, amsthm}
\usepackage{mathrsfs,cancel}
\usepackage{verbatim}
\usepackage{bm}
\usepackage{amssymb}
\usepackage{hyperref}
\usepackage{inputenc,array,subfig}
\usepackage{tikz}
\usetikzlibrary{decorations.markings}
\usepackage{mathtools}
\usepackage{textcomp}
\usepackage{appendix}
\usepackage{epsfig}
\usepackage{graphicx}
\allowdisplaybreaks
\usepackage{color}
\newcommand{\e}{\epsilon}
\newcommand{\be}[1]{\begin{equation}\label{#1} }
\newcommand{\ee}{\end{equation}}
\newcommand{\bea}[1]{\begin{eqnarray}\label{#1} }
\newcommand{\eea}{\end{eqnarray}}

\newcommand{\de}{\delta}

\renewcommand{\d}{\partial}

\newcommand{\D}{\Delta}

\renewcommand{\t}{\tau}
\newcommand{\s}{\sigma}
\definecolor{darkblue}{rgb}{0.0, 0.2, 0.6}

\newcommand{\nn}{\nonumber}

\definecolor{bluegray}{rgb}{0.2, 0.45, 0.8}

\title{Scattering of Null strings  -- Flipped Vacuum \& CHY}

\author[a]{Arjun Bagchi,} \author[a]{Sachin Grover,} \author[a]{Sharang Rajesh Iyer,} \author[a]{Amartya Saha,}
\author[b]{Stephan Stieberger.}
\author{\\}

\affiliation[a]{Indian Institute of Technology Kanpur, Kalyanpur, Kanpur, Uttar Pradesh 208016, India \\}
\affiliation[b]{Max-Planck Institut f\"ur Physik, Werner--Heisenberg--Institut, Boltzmannstr. 8, 85748 Garching bei M\"unchen, Germany\\}

\emailAdd{(abagchi, saching, siyer, amartyas)@iitk.ac.in, stieberg@mpp.mpg.de}
\abstract{We investigate the scattering of null tensionless strings. Classical null strings, given by the ILST action, give rise to three distinct quantum theories built on different vacua and representations of the underlying 2d Conformal Carroll algebra (CCA) which form the residual gauge symmetries on the null worldsheet. In this paper, we are interested in the quantum theory built out of the so-called ``flipped'' vacuum which realises the highest weight representation of the 2d CCA. By considering a novel class of vertex operators, intimately connected to the compactified null string, we build scattering amplitudes of the theory. We show that one naturally obtains the Cachazo-He-Yuan (CHY) formulae when one considers scattering of massless states in the null ``flipped'' string.}

\preprint{MPP-2026-152}
\begin{document}
\maketitle
\flushbottom

\section{Introduction} 
String theory is one of the most successful frameworks for constructing a consistent theory of quantum gravity. The length of the fundamental string $\ell_s = \sqrt{\alpha'}$ is the only tunable parameter of the free theory. This dictates the various regimes the theory can probe. $\alpha' \to 0$ is the limit where the fundamental string shrinks to a point, and one recovers (super) gravity from string theory. The limit $\alpha' \to \infty$, on the other hand, probes the very stringy regime and, in flat target spacetime, is connected to the tensionless limit of string theory where null strings  \cite{Schild:1976vq, Isberg:1993av}. 

\medskip

Scattering in string theory gets particularly interesting in the very high energy limit. Gross and Mende \cite{Gross:1987kza, Gross:1987ar} studied this ultra high energy (fixed-angle) limit of string scattering and found universal semi-classical behaviour very different from point-particle field theory results. In particular, they found that the fixed-angle string scattering was exponentially suppressed at high energies, in stark contrast to field-theory amplitudes, which fall off as power laws. This exponential softness was a hallmark of the extended nature of strings as opposed to point particles. 

\medskip

The Gross-Mende limit has historically always been examined as a limit on string scattering and not explored in terms of deformed worldsheet symmetries in the extreme high energy limit until recently. The tensionless null string provides a pathway to addressing this question through worldsheet methods, and that is what some of us attempted in \cite{Bagchi:2026iyu}. The null string has been examined in detail over the past decade in the so-called ILST formulation \cite{Isberg:1993av}. Some of the relevant literature is \cite{Bagchi:2013bga, Bagchi:2015nca, Bagchi:2016yyf, Casali:2016atr, Bagchi:2017cte, Casali:2017zkz, Bagchi:2018wsn, Bagchi:2019cay, Bagchi:2020ats, Bagchi:2020fpr, Chen:2023esw, Banerjee:2024fbi, Banerjee:2023ekd, Bagchi:2024qsb, Bagchi:2025jgu, Duary:2025hdb, Figueroa-OFarrill:2025njv, Figueroa-OFarrill:2026igk}. For further details and a more comprehensive list of references, we point the reader to the recent review \cite{Bagchi:2026wcu}. 

\subsection*{Carrollian Worldsheet Symmetries and Quantization}

The main feature that has triggered the recent advancements in the field is that the symmetries of the worldsheet morph from the usual two copies of the Virasoro algebra to the 2d conformal Carroll algebra (CCA$_2$) or the 3d Bondi-van der Burgh-Metzner-Sachs (BMS$_3$) algebra \cite{Barnich:2006av, Isberg:1993av, Bagchi:2013bga, Bagchi:2015nca}, the commutators of which are given by: 
\begin{subequations}\label{BMS3}
  \begin{align}
    & [L_n, L_m] = (n-m) L_{n+m} + \frac{c_L}{12}\delta_{n+m,0} (n^3-n), \\
    & [L_n, M_m] = (n-m) M_{n+m} + \frac{c_M}{12}\delta_{n+m,0} (n^3-n), \\
    & [M_n, M_m] = 0.
\end{align}  
\end{subequations}
In the above, $c_L, c_M$ are the central terms.  Carrollian and conformal Carrollian symmetries have appeared in a wide variety of contexts, including notably in holography of asymptotically flat spacetimes. The reader is referred to the comprehensive review \cite{Bagchi:2025vri} for an overview of recent developments and for a detailed list of references. 

\medskip

Carrollian symmetries appear in the vanishing speed of light ($c$) limit of Poincare symmetries, and similarly Carroll CFTs are the $c\to 0$ limit of relativistic CFTs. In the context of string theory, the string length plays the role of the effective speed of light and two copies of the Virasoro algebra ($\mathcal{L}^\pm_n$) on the worldsheet contract to \eqref{BMS3} \cite{Bagchi:2013bga, Bagchi:2015nca}:  
\begin{align}\label{eq: URcontract}
    L_n = \mathcal{L}_n^+ - \mathcal{L}_{-n}^-, \quad  M_n = \epsilon\left(\mathcal{L}_n^+ + \mathcal{L}_{-n}^-\right). 
\end{align}
where $\epsilon$ is the dimensionless parameter which we will take to zero in the limit. The use of methods of the CCA has been central to the recent explorations of tensionless null strings.

\medskip

One of the primary curiosities of the quantum null string was the discovery that a careful canonical quantization of the classical bosonic ILST string led to three inequivalent quantum theories depending on the choice of vacuum \cite{Bagchi:2020fpr}. These are 
\begin{itemize}
    \item Induced vacuum - based on the induced representations of \eqref{BMS3} \cite{Barnich:2014kra, Campoleoni:2016vsh}. 
    \item Flipped vacuum - based on the highest weight representations of \eqref{BMS3} \cite{Bagchi:2009ca}. 
    \item Oscillator vacuum - based on an intrinsic oscillator construction \cite{Bagchi:2015nca}. 
\end{itemize}
The examination of the Gross-Mende regime of the tensile theory in \cite{Bagchi:2026iyu} was based on the ILST string in the induced vacuum: 
\begin{align}
    M_0 |m,s\rangle = m |m,s\rangle,  \quad L_0 |m,s\rangle =  s |m,s\rangle, \quad   M_n |m,s\rangle =0,  \, \forall n \neq 0. 
\end{align}
Here, the states $|m,s\rangle$ are labelled with their $M_0, L_0$ eigenvalues $m, j$ and all other $M_n$ generators annihilate the state. This choice was made since this is the theory that one lands up in if one followed the limit \eqref{eq: URcontract}. This can be easily seen by following the Virasoro highest weight representations in the Carroll limit. In this case, it was important to stay just below the $\alpha'=\infty$ or exact tensionless or null point, since we were interested in the very high energy sector of usual string theory and not any peculiarities that may be associated with the exact null string \footnote{See \cite{Sheikh-Jabbari:2026vqh, Sheikh-Jabbari:2026tpf, Lindstrom:2026zno} for some further symmetries of the null ILST string at exactly the tensionless point that we don't consider here or in \cite{Bagchi:2026iyu}. If these symmetries are considered as gauge symmetries, the exact null spectrum may be different. But there seem to be issues with quantizating these theories with the gauged version of these symmetries \cite{Chen:2026cau, Duary:2026lmk}. These extra symmetries would not be relevant in our work.}. It was found that the ILST string in the induced vacuum naturally focussed on the high energy sector. We obtained the Gross-Mende saddle from a scattering calculation based on vertex operators defined on the Carrollian worldsheet and specialised to the induced vacuum \cite{Bagchi:2026iyu}. 

\subsection*{Theory on Flipped Vacuum}
In this paper, we focus on the flipped vacuum, which realises the highest weight representation of the BMS algebra: 
\begin{align}
    L_0 |\Delta, \xi\rangle = \Delta |\Delta, \xi\rangle,  \quad M_0 |\Delta, \xi\rangle =  \xi |\Delta, \xi\rangle, \quad   L_n |\Delta, \xi\rangle =0, \, M_n |\Delta, \xi\rangle =0,  \, \forall n > 0. 
\end{align}
The ILST string quantized in this vacuum has properties very different from what can be expected in a usual string theory, with only the massless sector of string states being allowed in an uncompactified theory. It has been argued that this could be derived from a ``twisted'' tensile string theory where one starts with the highest weight representation of left-moving Virasoro $\mathcal{L^+}$ and the lowest weight representation of right-moving Virasoro $\mathcal{L^-}$, and hence the name flipped. This ``twisted'' string theory also has a limited and only a massless spectrum \cite{Lee:2017utr}.\footnote{ When some of the spacetime directions are compactified, the spectrum is enhanced to an infinite number of higher spin excitations which may also be massive \cite{Casali:2017mss}.}

\medskip

Rather surprisingly, it was found that the null string in this flipped vacuum bore close similarities with the Ambitwistor string \cite{Casali:2016atr, Casali:2017zkz}. The ambitwistor string is a chiral worldsheet theory whose target space is the space of null geodesics \cite{Mason:2013sva}. This is a worldsheet theory designed to produce massless field theory amplitudes. The imposition of constraints on the ambitwistor worldsheet naturally leads to the Cachazo-He-Yuan (CHY) formula \cite{Cachazo:2013hca, Cachazo:2013gna} of massless scattering.

\medskip

The equivalence between the ambitwistor string and the ILST string was principally established in terms of the classical theory, where the ILST action was mapped to the ambitwistor action in \cite{Casali:2016atr, Casali:2017zkz}. It was argued that these theories are classically equivalent on-shell, whereas the quantum-mechanical theories may differ. Chirality of the null string was established using the highest weight representation and an analysis of null vectors following \cite{Bagchi:2009pe, Bagchi:2012yk}, which established that for $c_M=0$ and $\xi=0$, the algebra \eqref{BMS3} reduces to its Virasoro subalgebra. However, it was later understood (see e.g. \cite{Hao:2021urq}) that this is true for the so-called singlet representation and the algebra \eqref{BMS3} allows for more complicated multiplet representations mimicking logarithmic conformal field theories. Recent studies of the ILST string in this highest weight vacuum indicate that indeed the inequivalence of the null string and the bosonic ambitwistor string is plausible \cite{Figueroa-OFarrill:2025njv, Figueroa-OFarrill:2026igk} \footnote{We should emphasise, however, that these references \cite{Figueroa-OFarrill:2025njv, Figueroa-OFarrill:2026igk} refer to the formulation as the ``revisited'' bosonic ambitwistor instead of the ILST in the flipped vacuum.}. 

\subsection*{Our explorations in this paper}

In light of the above, it is important to ask how the scattering of the ILST string in the HW vacuum behaves and whether it truly encodes the field theoretic $\alpha' \to 0$ features in its intrinsic formulation. This is the question we pose in this paper. We will find that using a set of novel vertex operators designed for the HW vacuum, we are able to compute the n-point scattering amplitude of these ILST null strings. These vertex operators, first introduced in \cite{Bagchi:2026iyu}, generalise the ones obtained in the context of BMS scalars in \cite{Hao:2021urq}. Unlike the scattering in the induced vacuum, which involved a saddle-point analysis around the leading Gross-Mende saddle that survived the Carrollian limit on the worldsheet, the amplitude here is expressed as an integration that can be done {\it{exactly}}. Our intrinsic worldsheet analysis in the highest weight vacuum of the quantum ILST string leads to null string scattering amplitudes which directly encode the CHY formulae. 

\medskip

The vertex operators we use naturally lead us to describing the target space theory compactified on a circle {\footnote{Compactified ILST strings have been considered in detail in \cite{Banerjee:2023ekd, Banerjee:2024fbi}. What is fundamentally new in our work is the construction of vertex operators and, of course, the calculation of amplitudes from them.}}. This is reminiscent of the Gomis-Ooguri non-relativistic string \cite{Gomis:2000bd}, which also required a compact direction to set a scale in the problem. Unlike the Gomis-Ooguri string, however, the worldsheet here is not Lorentzian but a null surface and is therefore described by the conformal Carroll algebra rather than the two copies of the Virasoro. 

\medskip

We will see later in the paper that our graviton vertex operators lead to gravity scattering amplitudes that differ from those of the usual Einstein theory. One important question is thus the identification of the theory in the bulk that the propagating ILST string induces. We will report on this important issue and further emergent double copy structures in our companion paper \cite{Bagchi:2026pjx}. 

\medskip

The manuscript is organised as follows. In \cref{sec: The ILST action,sec: Canonical quantization}, we discuss the classical and quantum mechanical aspects of the ILST action, respectively. In \cref{sec: Correlation functions}, we analyse the two-point correlators of various fields in the flipped vacuum. Section \ref{sec: The vertex operators} focuses solely on the ``new" vertex operator and its characteristics. In \cref{sec: The physical fields}, we determine the physical fields up to level two and their properties. Next, in \cref{sec: Scattering amplitudes}, we compute the scattering amplitude of level 1 and level 2 physical states (focussing on vector bosons and gravitons) and demonstrate the factorisation property of the four-point graviton amplitude. Finally, we summarise the important aspects of our investigation and conclude in \cref{sec: summary}.

\medskip
  
The appendices are organised as follows. In \cref{App:CorrFn Details}, we give some more details on the computation of various correlation functions. In \cref{App:OPEs}, we summarise the OPEs of the physical fields up to level two with the components of the stress tensor. Appendix \ref{sec: B Field analysis} analyses the Kalb-Ramond field in some more detail.

\section{Classical Null Strings: The ILST action}\label{sec: The ILST action}

We begin with a recap of the classical null string. It is good to mention that when speaking of these strings, we will use the terminology null strings and tensionless strings interchangeably throughout the paper. We will specialise to a flat target spacetime, and hence null strings are tensionless here, and vice versa. When the spacetime has another intrinsic scale, like the radius of the spacetime for AdS spacetimes, strings can be tensionless without being null. We will not be interested in such cases and the strings we consider would naturally be propagating in flat backgrounds. 

\medskip

The ILST action for tensionless string propagating in flat spacetime is given as
\begin{equation}
    S_{\text{ILST}}=\frac{1}{2}\int d^2\sigma\,V^aV^b\,\d_aX^\mu\d_bX^\nu\,\eta_{\mu\nu}\,,
\end{equation}
where $\sigma^a=(\tau,\sigma)$ denotes coordinates, $V^a$ is a vector-density of weight $\frac{1}{2}$, and $X^\mu$ are scalar fields on the world-sheet. The classical equations of motion (e.o.m.) for $X^\mu$ following from the ILST action is
\begin{equation}\label{X-eom}
    \d_a\left(V^aV^b\,\d_bX^\mu\right)=0\,,
\end{equation}
subject to the following stationary-conditions of the variational method
\begin{equation}
\delta X^\mu(\tau_i,\sigma)=\delta X^\mu(\tau_f,\sigma)=0\hspace{5mm}\text{and}\hspace{5mm}
    \int\limits_{0}^{2\pi}d\s\,\d_{\s}\left(\delta X^\mu V^\sigma V^a\d_aX_\mu\right)(\tau,\sigma)=0\,.
\end{equation}
The e.o.m. for the non-dynamical field $V^a$ gives rise to the following constraints on the dynamics of $X^\mu$
\begin{align}\label{X-constraints}
    V^a\d_aX^\mu\d_bX_\mu=0\,.
\end{align}
Now, $\Gamma_{ab}=\d_aX^\mu\d_bX_\mu$ is the world-sheet metric induced from the target spacetime. Thus, \eqref{X-constraints} reveals that $V^a$ is an eigenvector of the induced metric $\Gamma_{ab}$ with zero eigenvalue. Hence, the tensionless world-sheet is a null surface.

\medskip

The canonical momenta conjugate to the scalars $X^\mu$ are
\begin{align}
    \Pi^{\mu}=V^\t V^a\d_aX^{\mu}\,,\text{\hspace{5mm} constrained as }\,\Pi^{\mu}\d_aX_\mu=0\,,
\end{align}
and the canonical stress tensor is
\begin{align}
    T^a_{\hspace{1.5mm}b}=V^aV^c\d_cX^\mu\d_bX_\mu-\frac{\delta^a_{\,b}}{2}V^cV^d\d_cX^\mu\d_dX_\mu\,.
\end{align}
We note that the canonical stress tensor is automatically traceless: $T^a_{\hspace{1.5mm}a}=0$. Furthermore, it is already conserved on the $X^\mu$-shell alone, i.e. $\text{\eqref{X-eom}}\Rightarrow\d_aT^a_{\hspace{1.5mm}b}=0$, and vanishes identically on the imposition of the $V^a$-shell constraints, i.e. $\text{\eqref{X-constraints}}\Rightarrow T^a_{\hspace{1.5mm}b}=0$. 

\medskip

We use worldsheet reparametrization invariance of the ILST action to choose the gauge: $V^a=\frac{1}{\sqrt{c'}}(1,0)$. From now on, we shall remain in this gauge unless stated otherwise. In this gauge, the e.o.m. and the constraints read
\begin{align}
    \text{e.o.m.:\hspace{5mm}}&\ddot X^\mu(\tau,\sigma)=0\,;\label{null-eom}\\
    \text{stationary conditions:\hspace{5mm}}&\delta X^{\mu}(\tau_i,\sigma)=\delta X^{\mu}(\tau_f,\sigma)=0\text{\hspace{2.5mm},\hspace{2.5mm}}\int\limits_{0}^{2\pi}d\s\,\d_\s\left(\dot X^\mu\delta X_{\mu}\right)(\tau,\sigma)=0\,;\nn\\
    \text{constraints:\hspace{5mm}}&\dot X^\mu\dot X_\mu=0=\dot X^\mu X'_\mu\,,\label{null-constraints}
\end{align}
while the conjugate momenta and the stress tensor are simplified as
\begin{align}
    \Pi^\mu=\frac{1}{c'}\dot X^\mu\hspace{5mm};\hspace{5mm}T^\t_{\hspace{1.5mm}\t}=\frac{1}{2c'}\dot X^\mu\dot X_\mu=-T^\sigma_{\hspace{1.5mm}\sigma}\hspace{2.5mm},\hspace{2.5mm}T^\t_{\hspace{1.5mm}\s}=\frac{1}{c'}\dot X^\mu X'_\mu\hspace{2.5mm},\hspace{2.5mm}T^\s_{\hspace{1.5mm}\t}=0\,.
\end{align}
After fixing the null-gauge, the ILST action enjoys the BMS$_3$ symmetry as the residual gauge symmetry; the corresponding world-sheet diffeomorphisms are
\begin{align}
    \s\rightarrow\s'=f(\s)\,\,\,,\,\,\t\rightarrow\t'=\t f'(\s)+g(\s)\,.
\end{align}

\medskip

There are two classes of solutions of the null-gauge e.o.m. \eqref{null-eom}, subjected to the periodic stationary condition $F^\mu(\t,\s)=F^\mu(\t,\s+2\pi)$ with $F\in\left\{\dot X\,,\de X\right\}$, that are relevant for string theory. We first note the time-dependent one\footnote{Closed strings cannot linearly depend on the $\sigma$ coordinate, unless the target spacetime has compactified directions. The $A_0$ mode is relevant when dealing with such spacetimes and should be ignored when the target spacetime has no compactified directions.}
\begin{align}
    X^\mu(\t,\s)=x^\mu+\sqrt{\frac{c'}{2}}A^\mu_0\s+\sqrt{\frac{c'}{2}}B^\mu_0\t+i\sqrt{\frac{c'}{2}}\sum\limits_{n\neq0} \frac{1}{n}\left[A^\mu_n-in\t B^\mu_n\right]e^{-in\s}\,,\label{Xexp}
\end{align}
that must be subjected to the constraints \eqref{null-constraints}. The time-independent solution, on the other hand, is
\begin{align*}
    Y^\mu(\t,\s)=y^\mu+\sqrt{\frac{c'}{2}}C^\mu_0\s+i\sqrt{\frac{c'}{2}}\sum\limits_{n\neq0} \frac{1}{n}\,C^\mu_n\,e^{-in\s}\,,
\end{align*}
which automatically satisfies \eqref{null-constraints}. In this work, we are mostly interested in the pair of solutions related to each other, as
\begin{align}
    \d_\t X^\mu(\t,\s)=\d_\s Y^\mu(\t,\s)\,,\label{XYrelation}
\end{align}
such that when the field $Y^\mu(\t,\s)$ has the following mode-expansion
\begin{equation}
Y^\mu(\t,\s)=y^\mu+\sqrt{\frac{c'}{2}}B^\mu_0\s+i\sqrt{\frac{c'}{2}}\sum\limits_{n\neq0} \frac{1}{n}\,B^\mu_n\,e^{-in\s}\,,\label{Yexp}
\end{equation}
the mode-expansion of $X^\mu(\t,\s)$ is given by \eqref{Xexp}. In terms of these modes, the components of the stress tensor are
\begin{align}
  T^\t_{\hspace{1.5mm}\t}(\t,\s)&=\frac{1}{4}\sum_{n\in\mathbb{Z}}\left(\sum_{m\in\mathbb{Z}}B^\mu_mB^\nu_{n-m}\,\eta_{\mu\nu}\right)e^{-in\s}\equiv\sum_{n\in\mathbb{Z}}M_n\,e^{-in\s}\,,\\
  T^\t_{\hspace{1.5mm}\s}(\t,\s)&=\frac{1}{2}\sum_{n\in\mathbb{Z}}\sum_{m\in\mathbb{Z}}\left(A^\mu_mB^\nu_{n-m}-in\frac{\t}{2} B^\mu_mB^\nu_{n-m}\right)\eta_{\mu\nu}\,e^{-in\s}\nn\\
  &\equiv\sum_{n\in\mathbb{Z}}\left(L_n-in\t M_n\right)e^{-in\s}\,.
\end{align}
In terms of the modes, the classical constraints on $X^\mu$ are $T^a_{\hspace{1.5mm}b}=0\Rightarrow L_n=0=M_n$.  

\medskip

The classical phase-space is governed by the following equal-time Poisson brackets
\begin{align}
    \left\{X^\mu(\t,\s_1)\,,X^\nu(\t,\s_2)\right\}=0\hspace{2.5mm},\hspace{2.5mm}&\left\{\Pi^\mu(\t,\s_1)\,,\Pi^\nu(\t,\s_2)\right\}=0\,,\nn\\
    \left\{X^\mu(\t,\s_1)\,,\Pi^\nu(\t,\s_2)\right\}&=2\pi\,\delta\left(\s_1-\s_2\right)\,\eta^{\mu\nu}\,.
\end{align}
These are translated into the following Poisson brackets between the modes appearing in \eqref{Xexp} and \eqref{Yexp}
\begin{align}
    &\left\{A^\mu_n\,,A^\nu_m\right\}=\left\{B^\mu_n\,,B^\nu_m\right\}=\left\{x^\mu\,,A^\nu_m\right\}=0\,,\nn\\
    &\left\{A^\mu_n\,,B^\nu_m\right\}=-2in\,\eta^{\mu\nu}\delta_{n+m}\hspace{2.5mm},\hspace{2.5mm}\left\{x^\mu\,,B^\nu_m\right\}={\sqrt{2c'}}\,\eta^{\mu\nu}{\delta_{m,0}}\,.\label{PB}
\end{align}
We note that the spacetime momentum $p^\mu$ of the tensionless string is given by 
\begin{equation}
    p^\mu=\int_{0}^{2\pi}\frac{d\s}{2\pi}\,\Pi^\mu=\frac{B^\mu_0}{\sqrt{2c'}}\,,
\end{equation}
and $\zeta^\mu:=\frac{A^\mu_0}{\sqrt{2c'}}$ characterizes the change in the target spacetime coordinate $X^\mu$ under a complete $\s$-cycle
\begin{equation}
    X^\mu(\t,\s+2\pi)-X^\mu(\t,\s)=\int_{0}^{2\pi}d\s\,X'^{\mu}=2\pi c'\zeta^\mu\,.\label{description A0}
\end{equation}

\medskip

Since the ILST action is invariant under a constant shift: $X^\mu\rightarrow X^\mu+a^\mu$, the mode-expansion \eqref{Xexp} allows for any $d$-dimensional toroidal compactification of the $D$-dimensional target spacetime (with $D>d$) \cite{Banerjee:2023ekd, Banerjee:2024fbi}. This is achieved by the following identification
\begin{align}
    X^a\equiv X^a+2\pi R^a\hspace{5mm}\text{for }\,D-d\leq a\leq D-1\,.\label{compact direction}
\end{align}
Furthermore, one can choose the tensionless string to wind $w(\in\mathbb{Z})$-times around a compact direction
\begin{align}
    X^a(\t,\s+2\pi)= X^a(\t,\s)+2\pi w^aR^a\,.\label{winding}
\end{align}

\section{Canonical quantization}\label{sec: Canonical quantization}
The classical tensionless string has three quantum avatars that pertain to different quantization conditions. The  Hilbert spaces corresponding to the three quantum tensionless string theories are built over the induced, flipped, and oscillator vacua \cite{Bagchi:2020fpr}. To quantize the ILST string, we impose the quantum constraints \eqref{null-constraints} on the physical states,
\begin{align}
\braket{\text{phys}\,|\,T^a_{~~b}\,|\,\text{phys}'}=0\,.\label{sandwich}
\end{align}
In terms of the modes of the stress tensor, we have the conditions,
\begin{equation}
\braket{\text{phys}\,|\,L_n\,|\,\text{phys}'}=\braket{\text{phys}\,|\,M_n\,|\,\text{phys}'}=0\, ,\quad n\neq 0\, .
\end{equation}
The physical states are eigenstates of the zero modes,
\begin{align}
   M_{0}\ket{\text{phys}}=a_M\ket{\text{phys}}\,,\quad L_{0}\ket{\text{phys}}=a_L\ket{\text{phys}}\,,\label{physstate-condition}
\end{align}
where $a_M$ and $a_L$ are the normal ordering constants. As stressed in the introduction and above, there are three distinct ways of quantizing the classical ILST string \cite{Bagchi:2020fpr}. The first option is 
\begin{equation}
    L_n\ket{\text{phys}}\neq 0\, ,\quad M_n\ket{\text{phys}}=0,\quad \forall n\neq0\, .
\end{equation}
This quantizes the system such that the states lie in the induced representation of the underlying BMS algebra. The second choice is  of the oscillator representation,
\begin{equation}
    L_n\ket{\text{phys}}\neq 0\, ,\quad M_n\ket{\text{phys}}\neq 0,\quad \forall n\in \mathbb{Z}\, ,
\end{equation}
but still \eqref{sandwich} is satisfied. This leads to an oscillator-like representation of the BMS algebra and hence the name. The third choice is of the flipped representation,
\begin{align}
   L_{n}\ket{\text{phys}}=0\,,\quad M_{n}\ket{\text{phys}}=0\,,\quad n>0\, ,\label{phystate-condition}
\end{align}
with $a_M=0\, ,a_L\neq 0$. This realises the highest weight BMS representations on the Hilbert space of states. We will focus our attention on this so-called flipped representation. 

\medskip

For canonical quantization, we use the Dirac rule on the classical Poisson brackets \eqref{PB} to obtain the following non-trivial commutators
\begin{align}
  \left[A^\mu_n\,,B^\nu_m\right]=2n\,\eta^{\mu\nu}\delta_{n+m}\,,\quad\left[x^\mu\,,B^\nu_m\right]=i{\sqrt{2c'}}\,\eta^{\mu\nu}{\delta_{m,0}}\,.  
\end{align}
Since $A^\mu_n$ and $B^\nu_m$ commute with each other when $n+m\neq0$, the quantum modes $L_{n}$ and $M_{m}$, for $n\neq0$ and $m\in\mathbb{Z}$, can be defined exactly the same way as in the classical case
\begin{equation}
    \begin{split}
        L_{n}=\frac{1}{2}\sum_{m\in\mathbb{Z}}\,A^\mu_mB^\nu_{n-m}\,\eta_{\mu\nu}\,,\quad M_{n}=\frac{1}{4}\sum_{m\in\mathbb{Z}}\,B^\mu_m B^\nu_{n-m}\,\eta_{\mu\nu}\,.\label{LM in AB}
    \end{split}
\end{equation}
However, the classical definition of the $L_0$-mode cannot be directly carried over to the quantum case due to the ordering ambiguity of the oscillator modes. The choice corresponding to the flipped vacuum,
\begin{align}
   A^\mu_{n}\ket{0}=0\,,\quad B^\mu_{n}\ket{0}=0\,,\quad \forall n>0,\label{AB annihilation}
\end{align}
implies a natural choice of ordering,
\begin{align}
   L_{0}= \frac{1}{2}A_0\cdot B_0+\frac{1}{2}\sum_{m=1}^\infty\left(  A_{-m}\cdot  B_{m}+  B_{-m}\cdot  A_{m}\right)\,:=\,\frac{1}{2}\sum_{m\in\mathbb{Z}}:  A_{-m}\cdot B_{m}:\,.
\end{align}
Furthermore, if we impose reality on $X^\mu$, we have the following hermitian conjugation relations
\begin{align}
  \left(  A^\mu_n\right)^\dagger=  A^\mu_{-n}\,,\quad\left(B^\mu_n\right)^\dagger= B^\mu_{-n}\quad\implies\quad L_{n}^\dagger=L_{-n}\,,\quad M_{n}^\dagger= M_{-n}\,.\label{hermitian conjugation}
\end{align}

\medskip

Since the hermitian operators $A^\mu_0$, $B^\mu_0$, $L_0$, and $M_0$ commute with each other, they should share a simultaneous eigenbasis. In this simultaneous eigenbasis, the $M_0$ ground state is completely specified as $\ket{0;p,\zeta}$ such that
\begin{equation}
    A^\mu_0\ket{0;p,\zeta}=\sqrt{2c'}\,\zeta^\mu\ket{0;p,\zeta}\,,\quad B^\mu_0\ket{0;p,\zeta}=\sqrt{2c'}\,p^\mu\ket{0;p,\zeta}\,.\label{zeromode annihilation}
\end{equation}
The target spacetime momenta are denoted as $p^a$ and $p^{\tilde\mu}$, where the index `$a$' runs along the compact directions \eqref{compact direction} and the index `$\tilde\mu$' runs along the non-compact directions. The compact momenta $p^a$ must be quantized as follows
\begin{align}
    p^a=\frac{n^a}{R^a}\,,\quad \forall \,n^a\in\mathbb{Z}\,.\label{p quantum}
\end{align}
Moreover, for a winding configuration \eqref{winding}, one immediately finds from \eqref{description A0} that $\zeta^\mu$ and $p^\mu\zeta^\mu$ (no sum over $\mu$) are also quantized \cite{Banerjee:2023ekd}
\begin{subequations}\label{zeta quantum}
\begin{gather}
c'\zeta^a=w^aR^a\quad \implies\quad c'\zeta^ap^a=n^aw^a\quad \forall a\in[D-d\,,\,D-1]\,,\label{eq: Winding number}\\
\zeta^{\tilde\mu}=0\quad \implies\quad c'\zeta^{\tilde\mu} p^{\tilde\mu}=0\quad \forall\tilde\mu\in[0\,,\,D-d)\,.
\end{gather}
\end{subequations}
The combination $c'p^\mu\zeta^\mu$ (no sum) is quantized to take integer values and is independent of the compactification radii.\footnote{The critical dimension of the null string in the flipped representation is $D=26$ \cite{Bagchi:2020fpr, Bagchi:2021rfw}.}

\medskip

We understand that $\ket{0;p,\zeta}=\ket{0}\otimes\ket{p,\zeta}$ is the true ground state of a single tensionless string with target spacetime momentum $p^\mu$ and another physical quantum number $\zeta^\mu$; the $\ket{0}$ part is only the vacuum on the world-sheet. Therefore, $\ket{0;0,\zeta}$ represents the family of infinitely degenerate Poincar\'e (target spacetime) invariant vacua. In view of the hermitian conjugation relations \eqref{hermitian conjugation}, $a_L\in\mathbb{R}$ captures the essence of the ordering ambiguity in the definition of $L_{0}$. However, requiring $a_L\neq0$, rules out the possibility that the states $\{\ket{0;0,\zeta}\}$ are physical states.

\medskip

From our definition of the vacuum, the following can be inferred
\begin{subequations}
   \begin{align}
    & L_{n}\ket{0;p,\zeta}= M_{n}\ket{0;p,\zeta}=0\,,\quad \forall n>0\,,\\
    & M_{0}\ket{0;p,\zeta}=\frac{c'p^2}{2}\ket{0;p,\zeta}\,,\quad M_{-1}\ket{0;p,\zeta}=\sqrt{\frac{c'}{2}}\,p_\mu B^\mu_{-1}\ket{0;p,\zeta}\\
    & L_{-1}\ket{0;p,\zeta}=\sqrt{\frac{c'}{2}}\left(p_\mu A^\mu_{-1}+\zeta_\mu B^\mu_{-1}\right)\ket{0;p,\zeta}\implies L_{-1}\ket{0;0,0}=0\,,\label{spacetransinv}\\
    & L_{0}\ket{0;p,\zeta}=c'p\cdot\zeta\ket{0;p,\zeta}\implies L_{0}\ket{0;0,\zeta}=0\ket{0;0,\zeta}\,.
\end{align} 
\end{subequations}
From the above, especially \eqref{spacetransinv}, it is clear that only the special vacuum $\ket{0;0,0}$ is invariant under the global $\text{iso}(2,1)$ part of the residual BMS$_3$ symmetry on the world-sheet.

\section{Correlation functions}\label{sec: Correlation functions}
We find the two-point correlation functions (Wightman) of the field $X^\mu$ on the vacuum $\ket{0;0,0}$ to be
\begin{align}
    \braket{X^\mu(\s_1,\t_1) X^\nu(\s_2,\t_2)}=-c'\eta^{\mu\nu}\left[\frac{i\t_1e^{i\s_1}-i\t_2e^{i\s_2}}{e^{i\s_1}-e^{i\s_2}}+i\pi\,\t_{12}\,\delta\left(\s_{12}\right)\right]\,.\label{XXWightCyl}
\end{align}
The retarded and advanced correlators are found from the Wightman function to be
\begin{align}\label{eq: ret/adv correlators}
    G^{\mu\nu}_{R/A}(\s_1,\t_1;\s_2,\t_2)&:=\pm\theta\left(\pm\t_{12}\right)\left[\left\langle X^\mu(\s_1,\t_1) X^\nu(\s_2,\t_2)\right\rangle-\left\langle X^\nu(\s_2,\t_2) X^\mu(\s_1,\t_1)\right\rangle\right]\nn\\
    &=\mp2\pi ic'\eta^{\mu\nu}\,\theta\left(\pm\t_{12}\right)\t_{12}\,\delta\left(\s_{12}\right)\,.
\end{align}
Finally, we have the following Feynman correlator (using $\theta(t)=\frac{1+\text{sgn}(t)}{2}$)
\begin{align}
    &G^{\mu\nu}_{F}(\s_1,\t_1;\s_2,\t_2):=\theta\left(\t_{12}\right)\left\langle X^\mu(\s_1,\t_1) X^\nu(\s_2,\t_2)\right\rangle+\theta\left(\t_{21}\right)\left\langle X^\nu(\s_2,\t_2) X^\mu(\s_1,\t_1)\right\rangle\nn\\
    =&\,-c'\eta^{\mu\nu}\left[\frac{i\t_1e^{i\s_1}-i\t_2e^{i\s_2}}{e^{i\s_1}-e^{i\s_2}}+i\pi\left|\t_{12}\right|\delta\left(\s_{12}\right)\right]\,.\label{XXFeynCyl}
\end{align}
The correlators $G^{\mu\nu}_{R/A/F}$ satisfy the two-point Schwinger-Dyson equation, while the Wightman functions \eqref{XXWightCyl} do not as expected,
\begin{align*}
  \d_{\t_1}^2 G^{\mu\nu}_{R/A/F}(\s_1,\t_1;\s_2,\t_2)=-2\pi ic'\eta^{\mu\nu}\delta\left(\t_{12}\right)\delta\left(\s_{12}\right)\,. 
\end{align*}

\medskip

The above Wightman functions and Feynman correlators are not $\t$-translation invariant. In contrast, the Feynman correlators are translation invariant on the $t$-$x$ plane, where we have the following BMS$_3$ cylinder $\rightarrow$ plane map
\begin{align}
    x=e^{i\s}\,\,\,,\,\,\,t=i\t e^{i\s}\,.\label{plane-cyl map}
\end{align}
In the plane coordinates, the above Feynman correlator reads, \footnote{The Euclidean correlator was also derived in \cite{Hao:2021urq}.}
\begin{align}
    G^{\mu\nu}_{F}(x_1,t_1;x_2,t_2)&=-c'\eta^{\mu\nu}\frac{t_{12}}{x_{12}}-\pi ic'\eta^{\mu\nu}\left|t_{12}\right|\delta\left(x_{12}\right)\,\nn\\
    &=-\lim\limits_{\e\to0^+}c'\eta^{\mu\nu}\frac{t_{12}}{x_{12}-i\e t_{12}}\,.\label{ieCarrFeyn}
\end{align}

Strictly speaking, to arrive at the plane correlator \eqref{ieCarrFeyn} from its cylinder counterpart \eqref{XXFeynCyl}, we treat $\left(x_i,t_i\right)$ as real variables and $\left(\s_i,\t_i\right)$ as their complex-valued functions given by the inverse of the map \eqref{plane-cyl map}, although we started with real coordinates $(\s,\t)$ on the world-sheet. This is \textit{a posteriori} justified, since the resulting expression \eqref{ieCarrFeyn} can be analytically continued to any complex $t_i$ and $x_i$ (such that $x_{12}\neq0$). From now on, unless otherwise specified, we shall perform a field theory analysis keeping both $x$ and $t$ real; if needed, the appropriate quantities will be analytically continued to complex $x$ and $t$.

\medskip

When $t$ is real, it is sensible to consider $t$-ordering. If we regard the plane Feynman correlator \eqref{ieCarrFeyn} to be $t$-ordered, we can read off the proper Wightman function on plane
\begin{gather}
    G^{\mu\nu}_{F}(x_1,t_1;x_2,t_2):=\theta\left(t_{12}\right)\braket{ X^\mu(x_1,t_1) X^\nu(x_2,t_2)}+\theta\left(t_{21}\right)\braket{ X^\nu(x_2,t_2) X^\mu(x_1,t_1)}\nn\\
    \implies\,\braket{ X^\mu(x_1,t_1) X^\nu(x_2,t_2)}=-c'\eta^{\mu\nu}\,t_{12}\left[\frac{1}{x_{12}}+i\pi\delta\left(x_{12}\right)\right]\,,\label{XXWightplane}
\end{gather}
which is consistent with the canonical commutation relation on the plane. For the sake of completeness, we also note the retarded/advanced correlators on the plane
\begin{align}
   G^{\mu\nu}_{R/A}(x_1,t_1;x_2,t_2)=\mp2\pi ic'\eta^{\mu\nu}\,\theta\left(\pm t_{12}\right)t_{12}\,\delta\left(x_{12}\right)\,. 
\end{align}

\medskip

We now proceed to find the Wightman function between the fields $X^\nu$ and $Y^\mu$, related by \eqref{XYrelation} or its plane version
\begin{align}
    \d_t X^\mu(t,x)=\d_x Y^\mu(t,x)\,.\label{XtoYplane}
\end{align}
Since time-ordering and time-differentiation do not commute with each other, it would be wise to impose the relation \eqref{XtoYplane} on the Wightman function \eqref{XXWightplane} to directly reach\footnote{Our branch cut convention on plane is \begin{align*}
  \lim\limits_{\e\to0} \log (x-i\e)=\log|x|-i\pi\,\theta(-x)\text{sgn}(\e)\text{\hspace{2.5mm} ; \hspace{2.5mm}}\log (x)=\log|x|-i\pi\,\theta(-x)\,. 
\end{align*}}
\begin{align}
    &\braket{\d_{x_1} Y^\mu(x_1,t_1) X^\nu(x_2,t_2)}=-c'\eta^{\mu\nu}\left[\frac{1}{x_{12}}+i\pi\delta\left(x_{12}\right)\right]\nn\\
   \implies\, &\braket{ Y^\mu(x_1,t_1) X^\nu(x_2,t_2)}=-c'\eta^{\mu\nu}\left[\log{\left|x_{12}\right|}-i{\pi}\,\theta\left(x_{21}\right)\right]=-c'\eta^{\mu\nu}\log{\left(x_{12}\right)}\,,\label{YX Wight}\\
   &\braket{ X^\nu(x_2,t_2)\d_{x_1} Y^\mu(x_1,t_1) }=-c'\eta^{\mu\nu}\left[\frac{1}{x_{12}}-i\pi\delta\left(x_{12}\right)\right]\nn\\
   \implies\, &\braket{ X^\nu(x_2,t_2) Y^\mu(x_1,t_1) }=-c'\eta^{\mu\nu}\left[\log{\left|x_{12}\right|}-i{\pi}\,\theta\left(x_{12}\right)\right]=-c'\eta^{\mu\nu}\log{\left( x_{21}\right)}\,,\label{XY Wight}
\end{align}
if we demand $t$-translation invariance. Hence, the $t$-ordered Feynman correlator between these two fields is
\begin{align}\label{eq: YY correlator}
    \braket{\hat{\mathcal{T}}\, Y^\mu(x_1,t_1) X^\nu(x_2,t_2)}
    &=-c'\eta^{\mu\nu}\left[\log{\left|x_{12}\right|}-i\pi\,\theta\left(-x_{12}t_{12}\right)\right]\\
    &=\lim\limits_{\e\to0^+}-c'\eta^{\mu\nu}\left[\log{\left(x_{12}-i\e t_{12}\right)}-i{\pi}\,\theta\left(t_{21}\right)\right]\,.\nn
\end{align}
Finally, by similar arguments and demanding $t$-translation invariance, we find that all kinds of two-point correlators of the $Y^\mu$ field vanish
\begin{align}
    \braket{Y^\mu(x_1,t_1) Y^\nu(x_2,t_2)}=0\,.\label{YYcorr}
\end{align}

\medskip

As discussed in Appendix \ref{App:CorrFn Details}, consistency of the Wightman functions \eqref{YX Wight}-\eqref{XY Wight} and \eqref{YYcorr} with the mode-expansions \eqref{Xexp} and \eqref{Yexp} demands
\begin{align}
    \left[y^\mu\,,B^\nu_n\right]=0\hspace{2.5mm},\hspace{2.5mm}\left[y^\mu\,,A^\nu_n\right]=i\sqrt{2c'}\,\eta^{\mu\nu}\delta_{n,0}\hspace{5mm}(\text{for }\, n\in\mathbb{Z})\,,
\end{align}
which reveals that $y^\mu$ and $\frac{A^\mu_0}{\sqrt{2c'}}$ form canonically conjugate pairs, the same as $x^\mu$ and $\frac{B^\mu_0}{\sqrt{2c'}}$.

\medskip

Since all of the above correlators on the vacuum $\ket{0;0,0}$ are translation invariant on the plane but not on the cylinder, we shall perform the world-sheet field theory analysis directly on the plane. We first translate the cylinder mode-expansions \eqref{Xexp} and \eqref{Yexp} of the fields $X^\mu$ and $Y^\mu$ into their (analytically continued) plane counterparts
\begin{subequations}
\begin{align}
    &X^\mu(t,x)=x^\mu-i\sqrt{\frac{c'}{2}}A^\mu_0\log x-i\sqrt{\frac{c'}{2}}B^\mu_0\frac{t}{x}+i\sqrt{\frac{c'}{2}}\sum\limits_{n\neq0} \frac{1}{n}\left[A^\mu_n-n\frac{t}{x} B^\mu_n\right]x^{-n}\,,\label{Xexp plane}\\
    &Y^\mu(t,x)=y^\mu-i\sqrt{\frac{c'}{2}}B^\mu_0\log x+i\sqrt{\frac{c'}{2}}\sum\limits_{n\neq0} \frac{1}{n}\,B^\mu_n\,x^{-n}\,.\label{Yexp plane}
\end{align}
\end{subequations}
Using these mode-expansions and \eqref{LM in AB}, the BMS$_3$ transformation properties of $X^\mu$ and $Y^\mu$ are found to be (with $n\in\mathbb{Z}$),
\begin{subequations}
\begin{align}
    &\left[M_n\,,Y^\mu(t,x)\right]=0\,,\quad\left[L_n\,,Y^\mu(t,x)\right]=x^{n+1}\d_x Y^\mu(t,x)\,,\label{Y trfo}\\
   &\left[M_n\,,X^\mu(t,x)\right]=x^{n+1}\d_t X^\mu(t,x)\,,\nn\\
&\left[L_n\,,X^\mu(t,x)\right]=\left[x^{n+1}\d_x+(n+1)x^nt\d_t\right]X^\mu(t,x)\,,\label{X trfo}
\end{align}
\end{subequations}
which resemble (for $n\geq-1$) those of a BMS$_3$ primary singlet field with $\Delta=0$ and $\xi=0$. However, only $Y^\mu$ can be considered a local primary field because its auto-correlator \eqref{YYcorr} is compatible with the Ward identities associated with its global BMS$_3$ transformation properties given by  \eqref{Y trfo} with $n=0,\pm1$.

\medskip

The two descendants $\d_xX^\mu$ and $\d_tX^\mu$($=\d_xY^\mu$) form a BMS$_3$ primary doublet with $\Delta=1$ and the $\bm{\xi}$-matrix \cite{Hao:2021urq, Bagchi:2026iyu},
\begin{equation}\label{eq: doublet}
    \bm{\xi}\begin{pmatrix}
        \d_xX^\mu\\
        \d_tX^\mu
    \end{pmatrix}=\begin{pmatrix}
        0 & 1\\
        0 & 0
    \end{pmatrix}\begin{pmatrix}
        \d_xX^\mu\\
        \d_tX^\mu
    \end{pmatrix}\,.\nn
\end{equation}
For later purposes, we list some non-zero two-point Feynman correlators, all of which can be obtained from the Wightman functions \eqref{XXWightplane}, \eqref{YX Wight}-\eqref{XY Wight}
\begin{subequations}
    \begin{align}
    \braket{\hat{\mathcal{T}}\d_{t_1}X^\mu(x_1,t_1)X^\nu(x_2,t_2)}&=\braket{\hat{\mathcal{T}}\d_{x_1}X^\mu(x_1,t_1)Y^\nu(x_2,t_2)}\nn\\&=\,-\lim\limits_{\e\to0^+}\frac{c'\eta^{\mu\nu}}{x_{12}-i\e t_{12}}\,,\label{X°XFeyn}\\
    \braket{\hat{\mathcal{T}}\d_{x_1}X^\mu(x_1,t_1)X^\nu(x_2,t_2)}&=\lim\limits_{\e\to0^+}{c'\eta^{\mu\nu}}\frac{t_{12}}{\left(x_{12}-i\e t_{12}\right)^2}\,,\label{X'XFeyn}\\
    \braket{\hat{\mathcal{T}}\d_{x_1}X^\mu(x_1,t_1)\d_{t_2}X^\nu(x_2,t_2)}&=\braket{\hat{\mathcal{T}}\d_{t_1}X^\mu(x_1,t_1)\d_{x_2}X^\nu(x_2,t_2)}\nn\\
    &=\,-\lim\limits_{\e\to0^+}\frac{{c'\eta^{\mu\nu}}}{\left(x_{12}-i\e t_{12}\right)^2}\,,\label{X'X°Feyn}\\
    \braket{\hat{\mathcal{T}}\d_{x_1}X^\mu(x_1,t_1)\d_{x_2}X^\nu(x_2,t_2)}&=\lim\limits_{\e\to0^+}{c'\eta^{\mu\nu}}\frac{2t_{12}}{\left(x_{12}-i\e t_{12}\right)^3}\,.\label{X'X'Feyn}
\end{align}
\end{subequations}
Comparison with the Feynman correlator \eqref{ieCarrFeyn} then reveals a useful trick: to obtain a Feynman correlator involving $\d_{t_{1}}X^\mu$ or $\d_{t_2}X^\mu$ directly from the corresponding one involving $X^\mu$, the combination $\left(x_{12}-i\e t_{12}\right)$ is effectively treated as time-independent. Since the correlators \eqref{X'X°Feyn}-\eqref{X'X'Feyn} along with $\braket{\hat{\mathcal{T}}\d_{t_1}X^\mu(x_1,t_1)\d_{t_2}X^\nu(x_2,t_2)}=0$ (and also the corresponding Wightman functions) are compatible with the appropriate global BMS$_3$ primary Ward identities, both $\d_tX^\mu$ and $\d_xX^\mu$ are local primary fields.

\section{Vertex operators}\label{sec: The vertex operators}
In addition to the above mentioned primary doublet \eqref{eq: doublet}, one can construct an infinite family of local primary singlets, the BMS vertex operators $\mathcal{V}_{p,\zeta}$, out of $X^\mu$ and $Y^\mu$ as below,
\begin{align}\label{genvertex}
    \mathcal{V}_{p,\zeta}(t,x)=\,:e^{ip\cdot X+i\zeta\cdot Y}:(t,x)\,,
\end{align}
with its $\Delta$ and $\xi$ both being real functions of $p^\mu$ and $\zeta^\mu$ given in \eqref{p quantum}-\eqref{zeta quantum}. In terms of the normal-ordered mode-expansion, we explicitly have
\begin{align}
    \mathcal{V}_{p,\zeta}(t,x)=\,e^{ip\cdot x+i\zeta\cdot y}\,&e^{-\sqrt{\frac{c'}{2}}\sum\limits_{n<0} \frac{1}{n}\left[p_\mu A^\mu_n+\left(\zeta_\mu-n\frac{t}{x}p_\mu\right) B^\mu_n\right]x^{-n}}\,e^{\sqrt{\frac{c'}{2}}\left[\left(p_\mu A^\mu_0+\zeta_\mu B^\mu_0\right)\log x+p_\mu B^\mu_0\,\frac{t}{x}\right]}\nn\\
   \times\, &e^{-\sqrt{\frac{c'}{2}}\sum\limits_{n>0} \frac{1}{n}\left[p_\mu A^\mu_n+\left(\zeta_\mu-n\frac{t}{x}p_\mu\right) B^\mu_n\right]x^{-n}}\,.\label{vexp}
\end{align}

\medskip

In order to demonstrate that the vertex operator \eqref{genvertex} is indeed a BMS$_3$ primary field, we first note the following commutators (for $n\in\mathbb{Z}$)
\begin{subequations}\label{AB-Vertex comm}
\begin{align}
    &\left[A^\mu_n\,,\mathcal{V}_{p,\zeta}(t,x)\right]=\sqrt{2c'}\left(\zeta^\mu+n\frac{t}{x}p^\mu\right)x^{n}\,\mathcal{V}_{p,\zeta}(t,x)\,,\\
    &\left[B^\mu_n\,,\mathcal{V}_{p,\zeta}(t,x)\right]=\sqrt{2c'}\,p^\mu x^{n}\,\mathcal{V}_{p,\zeta}(t,x)\,.
\end{align}
\end{subequations}
From these commutators, one finds that
\begin{subequations}\label{eigeneqn}
\begin{align}
   B^\mu_0\,\mathcal{V}_{p,\zeta}(t,x)\ket{0;0,0}&=\sqrt{2c'}\,p^\mu\,\mathcal{V}_{p,\zeta}(t,x)\ket{0;0,0}\,,\\
   A^\mu_0\,\mathcal{V}_{p,\zeta}(t,x)\ket{0;0,0}&=\sqrt{2c'}\,\zeta^\mu\,\mathcal{V}_{p,\zeta}(t,x)\ket{0;0,0}\,, 
\end{align}
\end{subequations}
i.e., the state $\mathcal{V}_{p,\zeta}(t,x)\vert0;0,0\rangle$ can describe a single string with a definite spacetime momentum $p^\mu$ and a winding quantum number $\zeta^\mu$. Using the same commutators, one obtains the following BMS$_3$ transformation properties of the vertex operator (for $n\in\mathbb{Z}$)
\begin{subequations}
\begin{align}
   & \left[M_n\,,\mathcal{V}_{p,\zeta}(t,x)\right]=\left[x^{n+1}\d_t+\frac{c'p^2}{2}(n+1)x^n\right]\mathcal{V}_{p,\zeta}(t,x)\label{MnVcomm}\,,\\
    & \left[L_n\,,\mathcal{V}_{p,\zeta}(t,x)\right]\nn\\
   &=\left[x^{n+1}\d_x+(n+1)x^{n}t\d_t+(n+1)x^n\,c'p\cdot\zeta+(n+1)nx^{n-1}t\,\frac{cp^2}{2}\right]\mathcal{V}_{p,\zeta}(t,x)\label{LnVcomm}\,,
\end{align}
\end{subequations}
which (especially for $n\geq-1$) confirms that $\mathcal{V}_{p,\zeta}$ possesses BMS$_3$ transformation properties appropriate for a primary singlet with
\begin{subequations}\label{eq: BMS3 labels Vop}
    \begin{align}
&\Delta=c'p\cdot\zeta=c'p_a\zeta^a=\sum_{a}n^aw^a\hspace{5mm}(\Delta\in\mathbb{Z})\,,\\ 
      &\xi=\frac{c'p^2}{2}=\frac{c'p_{\tilde\mu}p^{\tilde\mu}}{2}+\frac{c'}{2}\sum_{a}\left(\frac{n^a}{R^a}\right)^2\,.
    \end{align}
\end{subequations}
Following the methodology of \cite{Saha:2022gjw}, the commutators \eqref{MnVcomm}-\eqref{LnVcomm} can be obtained from the OPEs of the vertex operator with the BMS$_3$ stress tensor. The Carrollian OPE limit is achieved when the spatial separation between two operator insertions in an $m(\geq2)$-point Carrollian time-ordered Feynman correlator tends to (but does not exactly) vanish, regardless of their temporal separation \cite{Saha:2022gjw}. Since the two independent components of the quantum stress tensor are given on the plane by
\begin{align}
    T^t_{~~t}=-\frac{1}{2c'}:\d_t X^\mu\d_t X_\mu:\hspace{2.5mm},\hspace{2.5mm}T^t_{~~x}=-\frac{1}{c'}:\d_t X^\mu \d_xX_\mu:\hspace{2.5mm},\label{QMT}
\end{align}
we obtain the following OPEs between this stress tensor and the vertex operator \eqref{genvertex} (with $\lim~{\e\to0^+}$ implicit from now on)
\begin{subequations}
    \begin{align}
   T^t_{~~t}(1)\mathcal{V}_{p,\zeta}(2)\sim&\left[\frac{\frac{c'p^2}{2}}{\left(x_{12}-i\e t_{12}\right)^2}+\frac{\d_{t_2}}{x_{12}-i\e t_{12}}\right]\mathcal{V}_{p,\zeta}\left(2\right)\label{T1VOPE}\,,\\
   T^t_{~~x}(1)\mathcal{V}_{p,\zeta}(2)\sim&\left[\frac{{c'p\cdot\zeta}}{\left(x_{12}-i\e t_{12}\right)^2}+\frac{\d_{x_2}}{x_{12}-i\e t_{12}}\right.\nn\\
   &\left.\,-t_{12}\left\{\frac{{c'p^2}}{\left(x_{12}-i\e t_{12}\right)^3}+\frac{\d_{t_2}}{\left(x_{12}-i\e t_{12}\right)^2}\right\}\right]\mathcal{V}_{p,\zeta}\left(2\right)\label{T2VOPE}\,,
    \end{align}
\end{subequations}
which again confirms that the vertex operator is a BMS$_3$ primary singlet with $\Delta=c'p\cdot\zeta$ and $\xi=\frac{c'p^2}{2}$; here, $\sim$ denotes `up to terms non-singular in $x_{12}$'.

\medskip

By virtue of the mode-expansion \eqref{Xexp plane}, the (analytically continued) stress tensor components \eqref{QMT} are mode-expanded on plane as below
\begin{align}\label{Texp}
    T^t_{~~t}(t,x)=\sum_{n\in\mathbb{Z}}\,M_n\,x^{-n-2}\hspace{2.5mm},\hspace{2.5mm}T^t_{~~x}(t,x)=\sum_{n\in\mathbb{Z}}\left[L_n-(n+2)\frac{t}{x}M_n\right]x^{-n-2}\hspace{2.5mm}\,.
\end{align}
The OPEs between the stress tensor components and a general local Carrollian conformal field $\Phi$ are then expressed in the following forms \cite{Saha:2022gjw}
\begin{subequations}
\begin{align}
    &\,T^t_{~~t}(1){\Phi}(2)=\sum_{k=1}^n\frac{\left(M_{k-2}\Phi\right)(2)}{\left(x_{12}-i\e t_{12}\right)^{k}}+\sum_{k=0}^\infty{\left(x_{12}\right)^{k}}\,{\left(M_{-k-2}\Phi\right)(2)}\,,\\
    &\left[T^t_{~~x}-t_{12}\,\d_{x_1}T^t_{~~x}\right](1){\Phi}(2)=\sum_{k=1}^n\frac{\left(L_{k-2}\Phi\right)(2)}{\left(x_{12}-i\e t_{12}\right)^{k}}+\sum_{k=0}^\infty{\left(x_{12}\right)^{k}}\,{\left(L_{-k-2}\Phi\right)(2)}\,,
\end{align}
\end{subequations}
where the fields $\left(M_{-k-2}\Phi\right)$ and $\left(L_{-k-2}\Phi\right)$ with $k\geq-1$ are local descendants of the field $\Phi$. For the vertex operator $\mathcal{V}_{p,\zeta}$, we have the following descendants $(k\geq-1)$
\begin{subequations}\label{vertex descendants}
\begin{align}
    \left(M_{-k-2}\mathcal{V}_{p,\zeta}\right)&=\,\,:\left[\frac{ip\cdot\d_{x}^{k+1}\d_{t}X}{(k+1)!}+\frac{\d_{x}^k T^t_{\hspace{1.5mm}t}}{ k!}\right]\mathcal{V}_{p,\zeta}:\,\,,\\
    \left(L_{-k-2}\mathcal{V}_{p,\zeta}\right)&=\,\,:\left[\frac{\d_{x}^{k+2}(ip\cdot X+i\zeta\cdot Y)}{(k+1)!}+\frac{\d_{x}^kT^t_{\hspace{1.5mm}x}}{ k!}\right]\mathcal{V}_{p,\zeta}:\,\,.
\end{align}
\end{subequations}

\medskip

Finally, we note the $n$-point Wightman function and the Feynman correlator of the vertex operators
\begin{align}
   &\left\langle \prod_{i=1}^N \mathcal{V}_{p_i,\zeta_i}(t_i,x_i)\right\rangle_{x_i\neq x_j}=\,\delta^{(D-d)}\left(\sum_{i=1}^Np_i\right)\left[\,\prod_{a=D-d}^{D-1}\delta_{w^a_1+\ldots+w^a_N}\,\delta_{n^a_1+\ldots+n^a_N}\right]\nn\\
&\hspace{30mm}\times \exp\left[{c'\sum\limits_{i=1}^{N-1}\sum\limits_{j=i+1}^N\left(p_i\cdot p_j\,\frac{t_{ij}}{x_{ij}}+\left(p_i\cdot\zeta_j+p_j\cdot \zeta_i\right)\log\left(x_{ij}\right)\right)}\right]\,,\\
&\left\langle \hat{\mathcal{T}}\prod_{i=1}^N \mathcal{V}_{p_i,\zeta_i}(t_i,x_i)\right\rangle=\,\lim\limits_{\e\to0^+}\delta^{(D-d)}\left(\sum_{i=1}^Np_i\right)\left[\,\prod_{a=D-d}^{D-1}\delta_{w^a_1+\ldots+w^a_N}\,\delta_{n^a_1+\ldots+n^a_N}\right]\nn\\
 &\hspace{30mm}\times e^{c'\mathop{\sum\sum}\limits_{i<j}\left[p_i\cdot p_j\,\frac{t_{ij}}{x_{ij}-i\e t_{ij}}+\frac{p_i\cdot\zeta_j+p_j\cdot\zeta_i}{2}\left[\log\left(x_{ij}-i\e t_{ij}\right)+\log\left(x_{ji}-i\e t_{ji}\right)-i\pi\right]\right]}\,.\label{FeyncorrVertex}
\end{align}
It is straightforward to check that both the Wightman functions and the Feynman correlators of the vertex operators satisfy the appropriate Ward identities associated with their global BMS$_3$ transformation properties, establishing that the vertex operators \eqref{genvertex} are indeed local primary fields.

\section{Physical fields}\label{sec: The physical fields}
The main goal of this work is to compute the scattering amplitudes of tensionless bosonic strings with well-defined (target) spacetime momenta $\left\{p^\mu\right\}$ and internal quantum numbers $\left\{\zeta^\mu\right\}$. In view of \eqref{eigeneqn}, $\mathcal{V}_{p,\zeta}(t,x)|0;0,0\rangle$ may seem to describe the quantum state of such a single string. However, only the following state satisfies the $\{n>0\}$ physical state conditions \eqref{phystate-condition}, by virtue of the commutators \cref{MnVcomm,LnVcomm}
\begin{align}
    \lim\limits_{(t,x)\to(0,0)}\mathcal{V}_{p,\zeta}(t,x)|0;0,0\rangle\,.
\end{align}
Moreover, the $(n=0)$ physical state conditions are satisfied if
\begin{align}\label{on-shell condition for vertex}
    p^2=0\,\,\,\Longrightarrow\,\,\,-p_{\tilde\mu}p^{\tilde\mu}=\sum_{a}\left(\frac{n^a}{R^a}\right)^2\hspace{5mm}\,;\hspace{5mm}c'p\cdot\zeta=a_L\,\,\,\Longrightarrow\,\,\,a_L=\sum_an^aw^a\,.
\end{align}
However, $\mathcal{V}_{p,\zeta}$ is not the only local field that meets the above criteria.

\medskip

To construct other such local fields, we first consider the OPEs of the vertex operator $\mathcal{V}_{p,\zeta}$ with the primary doublet formed by $\d_tX^\mu$ and $\d_xX^\mu$
\begin{subequations}
\begin{align}
i\d_{t_1}X^\mu(1)\mathcal{V}_{p,\zeta}(2)=&\left[\frac{c'p^\mu}{x_{12}-i\epsilon t_{12}}\,\mathcal{V}_{p,\zeta}+\sum_{n=0}^\infty\frac{\left(x_{12}\right)^n}{n!}\,:i\d_{x_2}^n\d_{t_2}X^\mu\mathcal{V}_{p,\zeta}:\right](2),\\
i\d_{x_1}X^\mu(1)\mathcal{V}_{p,\zeta}(2)=&\left[\frac{c'\zeta^\mu}{x_{12}-i\epsilon t_{12}}\,\mathcal{V}_{p,\zeta}+\sum_{n=0}^\infty\frac{\left(x_{12}\right)^n}{n!}\,:i\d_{x_2}^{n+1}X^\mu\mathcal{V}_{p,\zeta}:\right.\nn\\
&\left.+t_{12}\left\{-\frac{c'p^\mu}{\left(x_{12}-i\epsilon t_{12}\right)^2}\,\mathcal{V}_{p,\zeta}+\sum_{n=0}^\infty\frac{\left(x_{12}\right)^n}{n!}\,:i\d_{x_2}^{n+1}\d_{t_2}X^\mu\mathcal{V}_{p,\zeta}:\right\}\right](2).
\end{align}
\end{subequations}
Denoting the OPEs of a general local field $\Phi$ with the said doublet as
\begin{align*}
    &\,i\d_{t_1}X^\mu(1){\Phi}(2)=\sum_{k=1}^m\frac{\left(B^\mu_{k-1}\Phi\right)(2)}{\left(x_{12}-i\e t_{12}\right)^{k}}+\sum_{k=0}^\infty\frac{\left(x_{12}\right)^{k}}{k!}\,{\left(B^\mu_{-k-1}\Phi\right)(2)}\,,\\
    &\left[i\d_{x_1}X^\mu-t_{12}\,\d_{x_1}i\d_{t_1}X^\mu\right](1){\Phi}(2)=\sum_{k=1}^n\frac{\left(A^\mu_{k-1}\Phi\right)(2)}{\left(x_{12}-i\e t_{12}\right)^{k}}+\sum_{k=0}^\infty\frac{\left(x_{12}\right)^{k}}{k!}\,{\left(A^\mu_{-k-1}\Phi\right)(2)}\,,
\end{align*}
we have, for the vertex operator, (with $k\geq0$)
\begin{subequations}
    \begin{align}
        &\left(B^\mu_{-k-1}\mathcal{V}_{p,\zeta}\right)=\,:i\d_{x}^k\d_{t}X^\mu\mathcal{V}_{p,\zeta}:\,,\\
    &\left(A^\mu_{-k-1}\mathcal{V}_{p,\zeta}\right)=\,:i\d_{x}^{k+1}X^\mu\mathcal{V}_{p,\zeta}:\,.
    \end{align}
\end{subequations}
Now, it can be shown using Wick's theorem that any local field of the following form (with all $k_i\,,l_j\geq0$)
\begin{align}
    \bm{\mathcal{V}}^{\mu_1\ldots\,\mu_n\,;\,\nu_1\ldots\,\nu_m}_{\bm p,\bm\zeta\,;\,k_1\ldots\,k_n\,;\,l_1\ldots\,l_m}\equiv&\left(A^{\mu_1}_{-k_1-1}\ldots\left(A^{\mu_n}_{-k_n-1}\left(B^{\nu_1}_{-l_1-1}\ldots\left(B^{\nu_m}_{-l_m-1}\mathcal{V}_{p,\zeta}\right)\right)\right)\right)\nn\\
    =&\,:i\d_x^{k_1+1}X^{\mu_1}\ldots\, i\d_x^{k_n+1}X^{\mu_n}\,i\d_x^{l_1}\d_tX^{\nu_1}\ldots\,i\d_x^{l_m}\d_tX^{\nu_m}\,\mathcal{V}_{p,\zeta}:\label{n level field}
\end{align}
satisfies \eqref{eigeneqn}. Each of the descendants \eqref{vertex descendants} of the vertex operator is generated by linear combinations of these fields, thus providing a good basis for expanding the OPEs in the world-sheet theory. However, if such a field $\bm{\mathcal{V}}^{\mu_1\ldots\,\nu_m}_{\bm p,\bm\zeta\,;\,k_1\ldots\,l_m}$ or its linear combination is not a primary, then $\lim\limits_{(t,x)\to(0,0)}\bm{\mathcal{V}}^{\mu_1\ldots\,\nu_m}_{\bm p,\bm\zeta\,;\,k_1\ldots\,l_m}(t,x)|0;0,0\rangle$ cannot satisfy all $(n>0)$ physical state conditions \eqref{phystate-condition}. So we shall look for such linear combinations of $\bm{\mathcal{V}}^{\mu_1\ldots\,\nu_m}_{\bm p,\bm\zeta\,;\,k_1\ldots\,l_m}$ that are primary fields.

\medskip
 
Since the scattering amplitudes of the tensionless string theory should be invariant under worldsheet diffeomorphisms, this implies, in particular that the integrated vertex operator insertions have to be diffeomorphism invariant. 
The integrated vertex operators are diffeomorphism invariant when the unintegrated local field $\bm{\mathcal{V}}^{\mu_1\ldots\,\nu_m}_{\bm p,\bm\zeta\,;\,k_1\ldots\,l_m}$ or its appropriate linear combination, whichever is a primary field, has scaling dimension $\Delta=2$ and is invariant under boost, $\xi=0$. The diffeomorphism invariance conditions are simply the physical state conditions \eqref{phystate-condition} with $a_L=2\, ,a_M=0$.

\medskip

Thus, the integrated version of the vertex operator $\mathcal{V}_{p,\zeta}$ will be world-sheet diffeomorphism invariant if we have
\begin{align}
    -p_{\tilde\mu}p^{\tilde\mu}=\sum_{a}\left(\frac{n^a}{R^a}\right)^2\hspace{5mm}\,;\hspace{5mm}\,\,\,\sum_an^aw^a=2\,\,,
\end{align}
as inferred from \eqref{on-shell condition for vertex}. Then, from the point of view of the physics on the non-compact dimensions, $\mathcal{V}_{p,\zeta}$ is a massive field with non-tachyonic $(\text{mass})^2$ $M^2=\sum\limits_{a}\left(\frac{n^a}{R^a}\right)^2>0\,$ when $\left\{R^a\right\}$ is finite. 

\medskip

To discuss the general case of the local field $\bm{\mathcal{V}}^{\mu_1\ldots\,\nu_m}_{\bm p,\bm\zeta\,;\,k_1\ldots\,l_m}$, we first note that
\begin{subequations}
    \begin{align}
    &\left[L_0\,,\,\bm{\mathcal{V}}^{\mu_1\ldots\,\nu_m}_{\bm p,\bm\zeta\,;\,k_1\ldots\,l_m}(t,x)\right]\nn\\
    &\,=\left[x\d_x+t\d_t+\left(c'p\cdot\zeta+\sum_{i=1}^n(k_i+1)+\sum_{j=1}^m(l_j+1)\right)\right]\bm{\mathcal{V}}^{\mu_1\ldots\,\nu_m}_{\bm p,\bm\zeta\,;\,k_1\ldots\,l_m}(t,x)\,,\\
      &\left[M_0\,,\,\bm{\mathcal{V}}^{\mu_1\ldots\,\nu_m}_{\bm p,\bm\zeta\,;\,k_1\ldots\,l_m}(t,x)\right]=\left[x\d_t+\frac{c'p^2}{2}\right]\bm{\mathcal{V}}^{\mu_1\ldots\,\nu_m}_{\bm p,\bm\zeta\,;\,k_1\ldots\,l_m}(t,x)\nn\\
    &\qquad\qquad\qquad\qquad\qquad\qquad\qquad+\sum_{i=1}^n(k_i+1)\,\bm{\mathcal{V}}^{\mu_1\ldots\,\mu_{i-1}\mu_{i+1}\ldots\,\mu_n\,;\,\mu_i\nu_1\ldots\,\nu_m}_{\bm p,\bm\zeta\,;\,k_1\ldots k_{i-1}k_{i+1}\ldots\,k_n\,;\,k_il_1\ldots\,l_m}(t,x)\,,\label{general boost matrix}
    \end{align}
\end{subequations}
which can be obtained just from the single and double poles (derived using Wick's theorem) in the OPEs of the said field with the stress tensor. We shall refer to the non-negative integral quantity $N\equiv\sum\limits_{i=1}^n(k_i+1)+\sum\limits_{j=1}^m(l_j+1)$ as the \textit{level}.

\subsection{Physical fields at level 1}
Having already discussed the level $0$ case, $\mathcal{V}_{p,\zeta}\,$, we now proceed to the next level where there are two local \textit{basis} fields --- $\,:i\d_{x}X^\mu\mathcal{V}_{p,\zeta}:$ and $:i\d_{t}X^\mu\mathcal{V}_{p,\zeta}:\,$. From \eqref{general boost matrix}, it is clear that any linear combination of these fields, which is boost invariant, must be of the form $:\e_\mu\d_{t}X^\mu\mathcal{V}_{p,\zeta}:\,$. As shown in Appendix \ref{App:OPEs}, this field is a primary singlet only when the condition $e\cdot p=0$ is satisfied. If this field is to generate a physical state, it must satisfy $c'p\cdot\zeta=1$ or $\sum\limits_an^aw^a=1$.

\medskip

Under the following infinitesimal shift of the polarization vector $\e_{\mu}$ of the primary level 1 operator,
\begin{align}
 \e_{\mu}\longrightarrow \e_{\mu}+\lambda(p)p_{\mu}\,, \label{vector gauge trfo}  
\end{align}
the level $1$ (unintegrated) physical field changes by a total derivative term
\begin{align*}
  :i\e_\mu\d_{t}X^\mu\mathcal{V}_{p,\zeta}:\,\,\longrightarrow\,\,:i\e_\mu\d_{t}X^\mu\mathcal{V}_{p,\zeta}:+\,\d_t\left(\lambda(p)\mathcal{V}_{p,\zeta}\right)\,,
\end{align*}
hence keeping its integrated version invariant. Moreover, the transformation \eqref{vector gauge trfo} preserves the primary field condition $\e\cdot p=0$. Therefore, \eqref{vector gauge trfo} describes a gauge transformation, and consequently, $:\e_\mu\d_{t}X^\mu\mathcal{V}_{p,\zeta}:$ represents a vector gauge boson with $(D-2)$ physical degrees of freedom (DoFs) in the $D$-dimensional target spacetime.

\medskip

From the perspective of \textit{non-compact} physics, we have one vector boson with $(D-d-1)$ DoFs and $(d-1)$ scalars as the complete set of physical fields, all of which are massive (for finite $\left\{R^a\right\}$) by virtue of the physical state condition $\sum\limits_an^aw^a=1$ and are not associated with any gauge symmetry in \textit{non-compact} physics.

\subsection{Physical fields at level 2}
Moving on to level $2$, we see from Appendix \ref{App:OPEs} that only the linear combinations summarised in \cref{tab:level2 physical states} can be physical primary fields which satisfy the primary state condition $p\cdot\zeta=0$.

\renewcommand{\arraystretch}{1.5}
\begin{table}[h!]
    \centering
    \begin{NiceTabular}{X[1,c,m]X[2,c,m]X[1,c,m]}[hvlines]
    Representation & State & Condition \\
     Symmetric tensor& $G_{\mu\nu}\left(\d_tX^\mu\d_tX^\nu+ic'p^{\mu}\,\d_t\d_xX^{\nu}\right)\mathcal{V}_{p,\zeta}$ & $p^{\mu}G_{\mu\nu}p^\nu=0$\\
     Antisymmetric tensor& $f_{\mu\nu}\d_tX^\mu\d_xX^\nu\,\mathcal{V}_{p,\zeta}$ & $f_{\mu\nu}p^\nu=p^{\mu}f_{\mu\nu}=0$\newline $f_{\mu\nu}\zeta^\nu=\zeta^{\mu}f_{\mu\nu}=0$\\
    \end{NiceTabular}
    \caption{Physical states at level 2}
    \label{tab:level2 physical states}
\end{table}

The following gauge transformation in the $D$-dimensional target spacetime
\begin{align}
 G_{\mu\nu}\longrightarrow G_{\mu\nu}+\lambda_{\mu}(p)p_{\nu}+\lambda_{\nu}(p)p_{\mu} \,\,\,\text{ with }\,\lambda\cdot p=0\,, \label{symm tensor gauge trfo}  
\end{align}
leaves the integrated version of the symmetric tensor boson (\cref{tab:level2 physical states}) invariant and also preserves the corresponding primary field condition: $p^{\mu}G_{\mu\nu}p^\nu=0$. The primary condition can be satisfied in two physically distinct ways.
\begin{enumerate}
    \item Let $G_{\mu\nu}=h_{\mu\nu}$ with $h_{\mu\nu}p^{\nu}=p^{\mu}h_{\mu\nu}=0$ and its gauge transformation given by \eqref{symm tensor gauge trfo}. The number of such independent tensors $h_{\mu\nu}$ is $\binom{D-1}{2}$.
    \item Let $V_{\mu}:=G_{\mu\nu}p^{\nu}\neq0$ such that $V\cdot p=0$. Since $V_{\mu}$ satisfies $V\cdot p=0$ and remains gauge-invariant under \eqref{symm tensor gauge trfo}, there are $(D-1)$ such independent $V_{\mu}$ (and the corresponding $G_{\mu\nu}$).
\end{enumerate}
The symmetric tensor $G_{\mu\nu}$ therefore contains $\binom{D}{2}$ physical DoFs in total. This counting of DoF can also be understood from the gauge transformation \eqref{symm tensor gauge trfo} and the associated primary condition.  

\medskip

\begin{table}[t]
\centering    
\renewcommand{\arraystretch}{1.5}
\begin{NiceTabular}[width=\textwidth]{X[1,c,m]X[2,c,m]X[1,c,m]X[2,c,m]}[hvlines]
     Particle  &  State & Condition & DOF\\
Graviton&$e_{\tilde{\mu}\tilde{\nu}}\d_tX^{\tilde{\mu}}\d_tX^{\tilde{\nu}}\,\mathcal{V}_{p,\zeta}$&$e_{\tilde{\mu}\tilde{\nu}}=e_{\tilde{\nu}\tilde{\mu}}$\newline $e_{~\tilde{\mu}}^{\tilde{\mu}}=0$\newline $e_{\tilde{\mu}\tilde{\nu}}p^{\tilde\mu}=0$\newline $e_{\tilde{\mu}\tilde{\nu}}p^{\tilde\nu}=0$& $\binom{D-d-1}{2}-1$\\
Dilaton&$\Pi_{\tilde\mu\tilde\nu}\d_tX^{\tilde{\mu}}\d_tX^{\tilde{\nu}}\,\mathcal{V}_{p,\zeta}$&-&1\\
Vectron&$q_{(\tilde{\mu}}v_{\tilde{\nu})}\left(\d_tX^{\tilde{\mu}}\d_tX^{\tilde{\nu}}\right.$ \newline $\left.~~~~~~+ic'p^{\tilde{\mu}}\,\d_t\d_xX^{\tilde{\nu}}\right)\mathcal{V}_{p,\zeta}$&$v_{\tilde{\mu}}p^{\tilde{\mu}}=0$& $D-d-2$\\
Scalaron&$q_{(\tilde{\mu}}p_{\tilde{\nu})}\left(\d_tX^{\tilde{\mu}}\d_tX^{\tilde{\nu}}\right.$\newline $\left.~~~~~~~~+ic'p^{\tilde{\mu}}\,\d_t\d_xX^{\tilde{\nu}}\right)\mathcal{V}_{p,\zeta}$&-&1\\
B-field&$f_{\tilde{\mu}\tilde{\nu}}\d_tX^{\tilde{\mu}}\d_xX^{\tilde{\nu}}\,\mathcal{V}_{p,\zeta}$&$f_{\tilde{\mu}\tilde{\nu}}=-f_{\tilde{\nu}\tilde{\mu}}$\newline $f_{\tilde{\mu}\tilde{\nu}}p^{\tilde\nu}=0$\newline $p^{\tilde\mu}f_{\tilde{\mu}\tilde{\nu}}=0$&$\zeta=0:\ \binom{D-d-2}{2}$\newline $
\zeta\neq0:\ \binom{D-d-1}{2}
$\\
Metric moduli&$\d_tX^{a}\d_tX^{b}\,\mathcal{V}_{p,\zeta}$&-&$\binom{d+1}{2}$\\
Vectron moduli&$\left(q_{\tilde{\mu}}\d_tX^{\tilde{\mu}}\d_tX^{a}\right.$\newline $\left.~~~~~~~~~~+\frac{ic'}{2}\,\d_t\d_xX^{a}\right)\mathcal{V}_{p,\zeta}$&-&$d$\\
B-field moduli&$f_{ab}\d_tX^{a}\d_xX^{b}\,\mathcal{V}_{p,\zeta}$&$f_{ab}=-f_{ba}$ \newline $f_{ab}\zeta^b=0$\newline $\zeta^af_{ab}=0$&$\zeta=0:\ 
\binom{d}{2}$\newline $
\zeta\neq0:\ \binom{d-1}{2}
$\\
KK gauge boson&$e_{\tilde{\mu}}\d_tX^{\tilde{\mu}}\d_tX^{a}\,\mathcal{V}_{p,\zeta}$&$e_{\tilde{\mu}}p^{\tilde{\mu}}=0$&$d(D-d-2)$\\
B-field gauge boson&$f_{\tilde{\mu}}F_a\left(\d_tX^{a}\d_xX^{\tilde{\mu}}\right.$\newline $\left.~~~~~~-\d_xX^{a}\d_tX^{\tilde{\mu}}\right)\mathcal{V}_{p,\zeta}$&$f_{\tilde{\mu}}p^{\tilde{\mu}}=0$\newline $F_a\zeta^a=0$&$\zeta=0:\ 
d(D-d-2)$\newline $\zeta\neq0:\
(d-1)(D-d-1)$\\
\end{NiceTabular}
\caption{Level 2 fields from the perspective of non-compact physics.}
\label{tab: Level 2 fields from non-compact physics}
\end{table}

\medskip

In contrast, the antisymmetric tensor boson (\cref{tab:level2 physical states}) transforms as a total derivative under the gauge transformation
\begin{align}
 f_{\mu\nu}\longrightarrow f_{\mu\nu}+\varepsilon_{\mu}(p)p_{\nu}-\varepsilon_{\nu}(p)p_{\mu}\,, \label{anti tensor gauge trfo}  
\end{align}
only if $\zeta^{\mu}=0$\,. Furthermore, to preserve the primary field condition, the gauge parameter must satisfy $\varepsilon\cdot p=0$. However, irrespective of the $\zeta^{\mu}$-data, the antisymmetric tensor $f_{\mu\nu}$ always contains $\binom{D-2}{2}$ physical DoFs. Therefore, the total number of physical DoFs that appear at level $2$ in the tensionless bosonic string theory is $(D-2)(D-1)+1$, as opposed to the $(D-2)^2$ massless DoFs in the usual closed tensile bosonic theory.

\medskip

The spectrum at level $2$ is special because only here it is possible to have massless physical fields even at finite $\left\{R^a\right\}$ (from the perspective of non-compact physics), thanks to the constraint $\sum\limits_an^aw^a=0$, which is satisfied by the choice: $\left\{n^a=0\Rightarrow p^a=0\right\}$. With this choice, the non-compact physics boasts of a variety of massless fields which are summarised in \cref{tab: Level 2 fields from non-compact physics}. The usual nomenclature from tensile string theory has been adopted (KK$\,:=\,$Kaluza-Klein), except the vectron, its moduli and the scalaron which altogether introduce the $(D-1)$ new physical DoFs exclusively in the tensionless theory.

\medskip

In \cref{tab: Level 2 fields from non-compact physics}, $e_{\tilde\mu\tilde\nu}$ and $v_{\tilde\mu}$ are defined as\footnote{Here $\Pi_{{\alpha}{\beta}}$ is the transverse projector for the null momentum $p^{\alpha}$, defined as: $\Pi_{{\alpha}{\beta}}:=\eta_{\alpha\beta}-p_{\alpha}q_{\beta}-p_{\beta}q_{\alpha}$ with $q^2=p^2=0$ and $p\cdot q=1$, where $q^{\alpha}$ is a reference null vector. For simplicity, we shall choose $\left\{q^a=0\right\}$.}
\begin{equation}
e_{\tilde{\mu}\tilde{\nu}}:=\left(\Pi_{\tilde{\mu}(\tilde{{\alpha}}}\Pi_{\tilde{\beta})\tilde{\nu}}-\frac{\Pi_{\tilde{\mu}\tilde{{\nu}}}\Pi_{\tilde{\alpha}\tilde{{\beta}}}}{D-d-2}\right)h^{\tilde{\alpha}\tilde{\beta}}\,,\quad v_{\tilde{\mu}}:=\Pi_{\tilde{\mu}\tilde{\nu}}V^{\tilde{\nu}}\,.
\end{equation}

\medskip

Evidently, the DoF associated with the antisymmetric tensor of the compactified parent theory fit into appropriate representations of $SO(D-d-2)\times SO(d)$ only in the special case $\zeta^a=0$. In the generic case, $\zeta^a\neq0$, these DoF are not associated with any gauge symmetry and furnish representations of $SO(D-d-1)\times SO(d-1)$. These massless physical fields surprisingly do not transform under the little group $SO(D-d-2)$ in the presence of a non-trivial winding. The underlying reason for this redistribution of the DoF associated with the two sectors of $\zeta^a$ requires further study, which we will not perform in this work.
The symmetric tensor boson, irrespective of the $\zeta^a-$ data, is a representation of $SO(D-d-2)\times SO(d)$. 

\medskip

When at least one $n^a$ is non-zero, each of the above mentioned fields acquires mass (from the perspective of non-compact physics) and consequently can no longer be associated with any non-compact gauge theory.

\section{Scattering amplitudes}\label{sec: Scattering amplitudes} 
In the usual tensile theory relation between the string scattering amplitudes and the correlators of the (integrated) vertex operators, the said correlator is Euclidean and radial-ordered (before integration). 

\medskip

In our case, we have noticed that only the time-ordered Feynman correlators are amenable to analytic continuation. To make a Feynman correlator Euclidean, we just analytically continue it to purely imaginary (Carrollian) time $t=it^E$, where $t^E\in\mathbb{R}$ is the Euclidean time. Under such analytic continuation, the $i\e$-prescription described above becomes redundant (when $x\in\mathbb{R}$) and can be omitted altogether. Then, the Euclidean version of the Feynman correlator \eqref{FeyncorrVertex} of the vertex operators is given as
\begin{align}\label{eq:EFcorrV}
 &\left\langle \hat{\mathcal{T}}\prod_{i=1}^N \mathcal{V}_{p_i,\zeta_i}(t_i,x_i)\right\rangle\xrightarrow[\text{continuation}]{\text{analytic}}\nn\\
 &\left\langle \prod_{i=1}^N \mathcal{V}_{p_i,\zeta_i}(t^E_i,x_i)\right\rangle=\delta^{(D-d)}\left(\sum_{i=1}^Np_i\right)\left[\,\prod_{a=D-d}^{D-1}\delta_{w^a_1+\ldots+w^a_N}\,\delta_{n^a_1+\ldots+n^a_N}\right]\nn\\
&\hspace{36.5mm}\times\exp\left[c'\mathop{\sum\limits_{i=1}^{N}\sum\limits_{j=1}^N}_{i\neq j}\left(it^E_i\,\frac{p_i\cdot p_j}{x_{ij}}+\frac{p_i\cdot\zeta_j+p_j\cdot\zeta_i}{4}\,\log\left( x^2_{ij}\right)\right)\right]\,.
\end{align}

\subsection{Symmetric tensors at level 2}\label{Graviton scattering amplitude}
We are interested mainly in the scattering of level $2$ fields, which we take to be massless (in \textit{non-compact} physics) by choosing $\left\{n^a=0\right\}$. This choice allows for the following simplifications
\begin{align*}
    p_i\cdot p_j:=p_{i\mu}p_j^{\mu}=p_{i\tilde{\mu}}p_j^{\tilde{\mu}}\hspace{5mm},\hspace{5mm}p_i\cdot \zeta_j=0\hspace{5mm},\hspace{5mm}G_{\mu\nu}^ip_j^{\mu}p_k^{\nu}=G_{\tilde{\mu}\tilde{\nu}}^ip_j^{\tilde{\mu}}p_k^{\tilde{\nu}}\,\,\,,\text{ etc}.
\end{align*}
This encourages us to further specialize to the case where all the winding numbers are fixed as $\left\{w^a=0\right\}$. 

\medskip

As a concrete example, let us focus on the symmetric tensor bosons (SBs), that is, the first four states in \cref{tab: Level 2 fields from non-compact physics}. The above choices simplify the corresponding vertex operators which can be collectively expressed as
\begin{align}
    \mathcal{V}^{\text{SB}}_p:=\,\,:G_{\tilde{\mu}\tilde{\nu}}\left(\d_tX^{\tilde{\mu}}\d_tX^{\tilde{\nu}}+ic'p^{\tilde{\mu}}\,\d_t\d_xX^{\tilde{\nu}}\right)e^{ip_{\tilde{\mu}}X^{\tilde{\mu}}}:\,\,.\label{non-compact Graviton Vertex no winding}
\end{align}
The tree-level scattering amplitude of $N$ SBs is then defined as
\begin{equation}\label{eq:n-point SB amplitude}
\mathcal{A}^{(N)}(p_1,\ldots,p_N):=\frac{(2\pi)^D\tilde{g}^N}{\text{Vol~(ISO(2,1))}}\int\prod_{i=1}^Ndx_idt_i^E\left\langle \prod_{i=1}^N \mathcal{V}^{\text{SB}}_{p_i}(t^E_i,x_i)\right\rangle\, .
\end{equation}

One can use the global subgroup $\text{ISO}(2,1)$ of the residual gauge symmetry to fix any three of the points of insertion. We choose to fix the following three points:
\begin{equation}\label{eq: 3-coord-fix}
    (t_1,x_1)=\lim\limits_{a\to\infty}(0,a)\,,~~~(t_2,x_2)=(0,1)\,,~~~(t_N,x_N)=(0,0)\,.
\end{equation}
The Faddeev-Popov determinant of this gauge fixing is found to be \cite{Casali:2017zkz}:
\begin{align}
    \D_{FP}=\left(x_{12}x_{2N}x_{N1}\right)^2=\lim\limits_{a\to\infty}a^4\,.
\end{align}

\subsection*{CHY amplitudes}
The $N$-point SB correlation function is obtained via Wick's theorem as
\begin{subequations}\label{eq graviton N point correlator}
\begin{align}
&\left\langle\prod_{i=1}^N\mathcal{V}_{p_i}^{\text{SB}}(t^E_i,x_i)\right\rangle=(ic')^{2N}\delta^{(D-d)}\left(\sum_{j=1}^Np_j\right)
\exp\left[c'\sum_{i=1}^{N-1}\sum_{j=i+1}^N\frac{it^E_{ij}}{x_{ij}}\,p_i\cdot p_j\right]\mathcal{I}^{\text{SB}}_N\,,\\
&\text{where}\hspace{2.5mm}\mathcal{I}^{\text{SB}}_N:=\prod_{i=1}^NG^{i}_{{{\mu}_i}{{\nu}}_i}\left[f_i^{{\mu}_i}f_i^{{\nu}_i}+p_i^{{\mu}_i}g_i^{{\nu}_i}\right]\hspace{2.5mm}\text{with}\hspace{2.5mm}f_i^{{\mu}} =\sum_{\substack{j=1\\j\neq i}}^N\frac{p_j^{{\mu}}}{x_{ij}}\hspace{2.5mm},\hspace{2.5mm}g_i^{{\mu}} =\sum_{\substack{j=1\\j\neq i}}^N\frac{p_j^{{\mu}}}{x^2_{ij}}\,\,,\label{eq f function}
\end{align}
\end{subequations}
which leads to the following $N$-SB amplitude from the definition \eqref{eq:n-point SB amplitude}
\begin{align}
    \mathcal{A}^{(N)}(p_1,\ldots,p_N)&=(2\pi)^D\tilde{g}^N\int\prod_{i=3}^{N-1}dx_idt_i^E\,\left(x_{12}x_{2N}x_{N1}\right)^2\left\langle \prod_{i=1}^N \mathcal{V}^{\text{SB}}_{p_i}(t^E_i,x_i)\right\rangle\nn\,\\
&\propto(-\tilde{g})^Nc'^{N+3}\left(x_{12}x_{2N}x_{N1}\right)^2\,\delta^{(D-d)}\left(\sum_{j=1}^Np_j\right)\nn\\
&\hspace{30mm}\times\int dx_3\ldots dx_{N-1}\,\mathcal{I}^{\text{SB}}_N\,\prod_{i=3}^{N-1}\delta\left(\sum\limits_{\substack{j=1\\j\neq i}}^N\frac{p_i\cdot p_j}{x_i-x_j}\right)\,.\label{Graviton CHY} 
\end{align}
The arguments of the delta-functions in the above integrand are known as the scattering equations \cite{Gross:1987ar}:
\begin{equation}
    \sum\limits_{\substack{j=1\\j\neq i}}^N\frac{p_i\cdot p_j}{x_i-x_j}=0\, , \quad 2< i< N\,.\label{Scatteringeqn}
\end{equation}
The form \eqref{Graviton CHY} makes it clear that the SB amplitude (built on the flipped vacuum) in the tensionless bosonic string theory belongs to the class of CHY constructible amplitudes; in this context, $\mathcal{I}^{\text{SB}}_N$ in \eqref{eq graviton N point correlator} may be called a CHY $N$-SB factor.\footnote{These amplitudes of the tensionless bosonic string are closely related to those obtained from the sectorised description of chiral strings. In this description of chiral strings, the integrated vertex operator is constructed using a sector-splitting operator \cite{LipinskiJusinskas:2019cej}. In the tensionless limit, the moduli space integration of the tree-level amplitude in the sectorised description localises onto the CHY scattering equations. The connection between chiral string amplitudes with CHY can also be established from Siegel's singular gauge choice \cite{Siegel:2015axg}.}

\subsection*{Two-point amplitude}
In tensile string theory, the constant factor in an on-shell two-point vertex operator correlator is proportional to the inner product of the corresponding one-particle state in the Hilbert space. Adopting the same interpretation, the norm of the one-SB state in the tensionless Hilbert space is seen to vanish \cite{Casali:2016atr, Banerjee:2023ekd}, as the two-point SB correlator \eqref{eq graviton N point correlator}$_{N=2}$ vanishes on-shell \footnote{Actually, for the graviton and the dilaton, the correlator \eqref{eq graviton N point correlator}$_{N=2}$ remains non-zero off-shell but vanishes on-shell due to transverse polarization; for the vectron and the scalaron, the two-point correlator already vanishes off-shell by virtue of momentum conservation.}. Since all of the on-shell two-point correlators vanish here, so do the $2$-SB amplitudes.

\subsection*{Three-point amplitude}
When $N=3$, there is no integral left to do in \eqref{Graviton CHY} since all three points of insertion have been gauge fixed, chosen to be \eqref{eq: 3-coord-fix}. One readily finds out that the $3$-point amplitude is non-zero
\begin{align}
    \mathcal{A}^{(3)}(p_1,p_2,p_3)
\propto&\,-\tilde{g}^3c'^6\left(x_{12}x_{23}x_{31}\right)^2\,\delta^{(D-d)}\left(\sum_{j=1}^3p_j\right)\mathcal{I}^{\text{SB}}_3\nn\\
=&\,\tilde{g}^3c'^6\,\delta^{(D-d)}\left(\sum_{j=1}^3p_j\right)\,G^1_{{\mu}{\nu}}\,p_2^{\mu}\,p_3^\nu\,G^2_{\rho\sigma}\,p_3^\rho\,p_1^\sigma\,G^3_{\alpha\beta}\,p_1^\alpha\,p_2^\beta\,.\label{Grav3}
\end{align}
Using momentum conservation and the on-shell conditions, it can be easily verified that this $3$-point amplitude is indeed gauge invariant on the worldsheet, i.e., independent of the choice \eqref{eq: 3-coord-fix}.

\subsection*{Four-point amplitude}
When $N=4$, the integrand in \eqref{Graviton CHY} is supported at the solution of the only scattering equation
\begin{equation}\label{eq: scattering-eqn}
   \sum\limits_{{j=1,2,4}}\frac{p_3\cdot p_j}{x_{3}-x_{j}}=0\hspace{2.5mm}\xRightarrow[\eqref{eq: 3-coord-fix}]{\text{gauge}}\hspace{2.5mm}x_3=-\frac{s}{t}\,.
\end{equation}
Consequently, the manifestly crossing symmetric $4$ point SB amplitude is obtained as
\begin{align}
   \mathcal{A}^{(4)}\propto&\,{\tilde{g}^4}{c'^7}\left(x_{12}x_{24}x_{41}\right)^2\,\delta^{(D-d)}\left(\sum_{j=1}^4p_j\right)\int dx_3\,\mathcal{I}^{\text{SB}}_4\,\delta\left(\sum\limits_{\substack{j=1,2,4}}\frac{p_3\cdot p_j}{x_3-x_j}\right)\nn\\
   =&\,\frac{\tilde{g}^4c'^7}{(stu)^3}\,\,\delta^{(D-d)}\left(\sum_{j=1}^4p_j\right)\,G^1_{\mu\nu}\left[\left(tp_2^\mu-sp_3^\mu\right)\left(tp_2^\nu-sp_3^\nu\right)+p_1^{\mu}\left(t^2p_2^\nu+s^2p_3^\nu\right)\right]\nn\\
&\times G^2_{\rho\sigma}\left[\left(tp_3^\rho-up_4^\rho\right)\left(tp_3^\sigma-up_4^\sigma\right)+p_2^{\rho}\left(t^2p_3^\s+u^2p_4^\s\right)\right]\,G^3_{\alpha\beta}\left[\left(sp_2^\alpha-up_4^\alpha\right)\left(sp_2^\beta-up_4^\beta\right)\right.\nn\\
&\left.+\,p_3^{\alpha}\left(s^2p_2^\beta+u^2p_4^\beta\right)\right]\,G^4_{\gamma\delta}\left[\left(sp_2^\gamma-tp_3^\gamma\right)\left(sp_2^\delta-tp_3^\delta\right)+p_4^{\gamma}\left(s^2p_2^\delta+t^2p_3^\delta\right)\right]\label{4 SB amp}\,.
\end{align}

\medskip

From the above $4$-SB amplitude, one can easily find any $4$-``particle'' amplitude involving the corresponding vertex operators just by substituting $G_{\mu\nu}$ with appropriate polarizations. For example, the $4$-graviton amplitude is found to be
\begin{align}
   \mathcal{A}^{(4)}\propto&\,\frac{\tilde{g}^4c'^7}{(stu)^3}\,\,\delta^{(D-d)}\left(\sum_{j=1}^4p_j\right)\,e^1_{\mu\nu}\left(tp_2^\mu-sp_3^\mu\right)\left(tp_2^\nu-sp_3^\nu\right)
\,e^2_{\rho\sigma}\left(tp_3^\rho-up_4^\rho\right)\left(tp_3^\sigma-up_4^\sigma\right)\nn\\
&\times e^3_{\alpha\beta}\left(sp_2^\alpha-up_4^\alpha\right)\left(sp_2^\beta-up_4^\beta\right)
\,e^4_{\gamma\delta}\left(sp_2^\gamma-tp_3^\gamma\right)\left(sp_2^\delta-tp_3^\delta\right)\,,\label{4 graviton amplitude}
\end{align}
upon imposing the transverse polarization condition. Clearly, the above amplitude is different from the $4$-graviton amplitude arising in Einstein gravity. 

\subsection*{Factorization of the graviton amplitude}
We now analyze the factorization of the $4$-graviton amplitude \eqref{4 graviton amplitude} at the cubic pole at $s=0$ (with $t$ fixed and $u=-t-s$). As $s\to0$, the said amplitude diverges as:
\begin{equation}
    \mathcal A^{(4)}\sim-\frac{\tilde g^4c'^7t^2}{s^3}\,e^1_{\mu\nu}\,p_2^\mu p_2^\nu\,e^2_{\rho\sigma}\,p_1^\rho p_1^\sigma\,e^3_{\alpha\beta}\,p_4^\alpha p_4^\beta\,e^4_{\gamma\delta}p_3^\gamma p_3^\delta\,=-\frac{\tilde g^4c'^7t^2}{s^3}\mathcal{P}\,.
\end{equation}
 We want to check whether the coefficient of the cubic pole factorizes as
\begin{equation}
    \lim_{s\to 0}s^3\mathcal A^{(4)}=\sum_I\sum_{\text{polarisation}_I}\mathcal A_{GGI}^{(3)}(1,2,k)\mathcal A_{IGG}^{(3)}(-k,3,4)\,,
\end{equation}
where $I$ denotes an exchanged intermediate particle and $k=-p_1-p_2=p_3+p_4$ is on-shell. 

\medskip

Since we have chosen the graviton to have no non-trivial winding, the exchanged particles $\{I\}$ must also have zero windings, by virtue of the winding number conservation in the three-point functions. Thus, in this case, only the level 2 particles with zero winding can be exchanged. Therefore, for ease of discussion, we can further take the target spacetime non-compact in all directions.

\medskip

\emph{Graviton exchange:} Let $e^k$ be the intermediate graviton polarisation. Using momentum conservation and transversality, the three-point graviton amplitude $\mathcal A^{(3)}(1,2,k)$ can be rewritten as 
\begin{equation}
\mathcal{A}_{GGG}^{(3)}(1,2,k)=\tilde{g}^3c'^6e^1_{\mu\nu}\,p_2^\mu\,p_2^\nu\,e^2_{\rho\sigma}\,p_1^\rho\,p_1^\sigma\,e^k_{\vartheta\varphi}\,p_1^\vartheta\,p_1^\varphi\,.
\end{equation}
Similarly, $\mathcal A_{GGG}^{(3)}(-k,3,4)$ is given as 
\begin{equation}
    \mathcal A_{GGG}^{(3)}(-k,3,4)=\tilde{g}^3c'^6\,e^{k}_{\lambda\kappa}\,p_3^\lambda\,p_3^\kappa\,e^3_{\alpha\beta}\,p_4^\alpha\,p_4^\beta\,e^4_{\gamma\delta}\,p_3^\gamma\,p_3^\delta\,.
\end{equation}
The product of three point amplitudes with the sum over the polarisations is given as 
\begin{equation}
\sum_{\text{pol}}\mathcal A_{GGG}^{(3)}(1,2,k)\mathcal A_{GGG}^{(3)}(-k,3,4)=\tilde{g}^6c'^{12}\mathcal{P}\sum_{\text{pol. }k}e^k_{\vartheta\varphi}\,e^{k}_{\lambda\kappa}\,p_1^\vartheta\,p_1^\varphi\,p_3^\lambda\,p_3^\kappa\,.
\end{equation}
Now, the sum over the polarisations is the spin-2 transverse traceless projector:
\begin{gather}
    \sum_{\text{pol. }k}e^k_{\vartheta\varphi}\,e^{k}_{\lambda\kappa}=\frac{1}{2}\left(\Pi_{\vartheta\lambda}\Pi_{\varphi\kappa}+\Pi_{\vartheta\kappa}\Pi_{\varphi\lambda}\right)-\frac{1}{D-2}\Pi_{\vartheta\varphi}\Pi_{\lambda\kappa}\,.
\end{gather}
Using this polarization sum, we finally obtain
 \begin{equation}
\sum_{\text{pol}}\mathcal A_{GGG}^{(3)}(1,2,k)\mathcal A_{GGG}^{(3)}(-k,3,4)=\frac{\tilde{g}^6c'^{12}t^2}{4}\mathcal{P}\,. 
\end{equation}
Hence, the four-graviton amplitude factorises (at the leading order cubic pole) to a product of two three-graviton amplitudes up to a multiplicative constant.

\medskip

\emph{Dilaton exchange:} we similarly analyse whether a dilaton is exchanged in the scattering process. The graviton-graviton-dilaton amplitude is given as 
\begin{equation}
    \mathcal{A}_{GGD}=\tilde{g}^3c'^6\,\delta^{(D)}\left(\sum_{j=1}^3p_j\right)\,e^1_{\mu\nu}\,p_2^\mu\,p_2^\nu\,e^2_{\rho\sigma}\,p_3^\rho\,p_3^\sigma\,\e^3_{\alpha\beta}\,p_1^\alpha\,p_1^\beta\,,
\end{equation}
where $\e_{\alpha\beta}$ is the transverse polarisation for the dilaton. The dilaton polarization sum is given as 
\begin{equation}
\sum_{\text{pol}}\e_{\mu\nu}^{(k)}\e^{(k)}_{\rho\sigma}=\frac{1}{D-2}\Pi_{\mu\nu}\Pi_{\rho\sigma}\,.
\end{equation}
The product of three point amplitudes with the sum over the polarisations is then given as 
\begin{equation}
\sum_{\text{pol}}\mathcal A_{GGD}^{(3)}(1,2,k)\mathcal A_{GGD}^{(3)}(-k,3,4)=\tilde{g}^6c'^{12}\mathcal{P}\sum_{\text{pol}}\e^k_{\vartheta\varphi}\,\e^{k}_{\lambda\kappa}\,p_1^\vartheta\,p_1^\varphi\,p_3^\lambda\,p_3^\kappa\,=0.
\end{equation}
Therefore, a dilaton-exchange does not contribute to the leading order singularity of the 4-graviton amplitude. 

\medskip

\emph{B-Field exchange:} The graviton-graviton-B field amplitude is given in \cref{eq: GGB amplitude}. Using the B-field polarization sum:
\begin{equation}
\sum_{\text{pol}}f_{\vartheta\varphi}^{(k)}f^{(k)}_{\lambda\kappa}=\frac{1}{2}\left(\Pi_{\vartheta\lambda}\Pi_{\varphi\kappa}-\Pi_{\vartheta\kappa}\Pi_{\varphi\lambda}\right)\,,
\end{equation}
The product of three point amplitudes with the sum over the polarisations is then given as 
\begin{align}
\sum_{\text{pol}}\mathcal A_{GGB}^{(3)}(1,2,k)\mathcal A_{GGB}^{(3)}(-k,3,4)
&=4c'^{10}e^1_{\mu\nu}\,p_2^\mu\,e^2_{\rho\sigma}\,p_1^\rho\,e^3_{\alpha\beta}\,p_4^\alpha\,e^4_{\gamma\delta}\,p_3^\gamma\\
&\times\left[p_1^\sigma\left(p_3^\delta\Lambda^{\beta\nu}-p_4^\beta\Lambda^{\delta\nu}\right)-p_2^\nu\left(p_3^\delta\Lambda^{\beta\sigma}-p_4^\beta\Lambda^{\delta\sigma}\right)\right]\nn\,,
\end{align}
where $\Lambda^{\mu\nu}$ is defined as 
\begin{equation}
    \Lambda^{\mu\nu}=\frac{1}{2}\left[p_2\cdot p_4\Pi^{\mu\nu}-\left(p_2^\mu-k^\mu(q\cdot p_2)\right)\left(p_4^\nu-k^\nu(q\cdot p_4)\right)\right]\,.
\end{equation}
This analysis shows that a B field-exchange also does not contribute at the leading order. 

\medskip

Therefore, at $\mathcal{O}\left(\frac{1}{s^3}\right)$, only the graviton contributes. From this analysis, we infer that the graviton propagator goes as $\frac{1}{p^6}$.  

\subsection{Vector at level 1}
As discussed before, the vector boson at level 1 must have non-trivial winding and hence, unlike those at level 2, can only become physical fields when at least one direction in the target spacetime is compact. We shall now turn to scattering amplitudes of these vector bosons from the point of view of the parent target spacetime.

\medskip

The tree-level scattering amplitude of $N$ vector bosons is similarly defined as,
\begin{align}\label{eq:n-point amplitude}
&\mathcal{A}^{(N)}(p_1,\ldots,p_N):=\frac{(2\pi)^D\tilde{g}^N}{\text{Vol~(ISO(2,1))}}\int\prod_{i=1}^Ndx_idt_i^E\left\langle \prod_{i=1}^N \mathcal{V}^{\text{V}}_{p_i,\zeta_i}(t^E_i,x_i)\right\rangle\,,\\
\text{with }\,\,\,&  \mathcal{V}^{\text{V}}_{p,\zeta}:=\,\,:e_{\nu}\,\d_t X^\nu\,e^{ip_{{\mu}}X^{{\mu}}+i\zeta_\mu Y^{\mu}}:\,\,.\nn
\end{align}
The $N$-point vector boson correlator is easily obtained as 
\begin{align}
    \left\langle{\prod_{i=1}^{N}\mathcal{V}^{\text{V}}_{p_i,\zeta_i}(t_i^E,x_i)}\right\rangle=\,&(-ic')^N\,\delta^{(D-d)}\left(\sum_{i=1}^Np_i\right)\left[\,\prod_{a=D-d}^{D-1}\delta_{w^a_1+\ldots+w^a_N}\,\delta_{n^a_1+\ldots+n^a_N}\right]\nn\\
&\times\exp\left[ic'\mathop{\sum\limits_{i=1}^{N-1}\sum\limits_{j=i+1}^N}t^E_{ij}\,\frac{p_i\cdot p_j}{x_{ij}}\right]\mathcal{I}_N^{\text{V}}\,,\label{corrE20}\\
\text{where }\,\,\mathcal{I}_N^{\text{V}}:=\,&\exp\left[c'\mathop{\sum\limits_{i=1}^{N-1}\sum\limits_{j=i+1}^N}p_{(i}\cdot\zeta_{j)}\log\left( x^2_{ij}\right)\right]\prod_{i=1}^Ne^i_{\mu_i}f^{\mu_i}_i\,,
    \nn
\end{align}
with $f^{\mu_i}_i$ defined in \eqref{eq f function}. Fixing three points of insertion using the worldsheet global residual gauge symmetry, the definition \eqref{eq:n-point amplitude} leads to the following CHY form of the $N$-point vector boson amplitude:
\begin{align}
    \mathcal{A}^{(N)}(p_1,\ldots,p_N)\propto&\,(-i\tilde{g})^Nc'^{3}\left(x_{12}x_{2N}x_{N1}\right)^2\,\left[\,\prod_{a=D-d}^{D-1}\delta_{w^a_1+\ldots+w^a_N}\,\delta_{n^a_1+\ldots+n^a_N}\right]\nn\\
&\times\,\delta^{(D-d)}\left(\sum_{j=1}^Np_j\right)\int dx_3\ldots dx_{N-1}\,\mathcal{I}^{\text{V}}_N\,\prod_{i=3}^{N-1}\delta\left(\sum\limits_{\substack{j=1\\j\neq i}}^N\frac{p_i\cdot p_j}{x_i-x_j}\right)\,.\label{Vector CHY}
\end{align}

\medskip

From \eqref{corrE20}$_{N=2}$, it is evident that the on-shell two-point correlator, and hence the two-point amplitude, of the vector boson vanishes, which is consistent with the computation of the norm of the level 1 state in \cite{Banerjee:2023ekd}.

\medskip

For the three-point function, after fixing the three points of insertion, there is no integration left in \eqref{Vector CHY}; we readily get,
\begin{align}
    \mathcal{A}^{(3)}(p_1,\ldots,p_3)\propto&\,(i\tilde{g}c')^3\left(e_1\cdot p_2\right)\left(e_2\cdot p_3\right) \left(e_3\cdot p_1\right)\label{Null3pt}\\
    &\times\,\delta^{(D-d)}\left(\sum_{i=1}^3 p_i\right)\left[\,\prod_{a=D-d}^{D-1}\delta_{w^a_1+\ldots+w^a_3}\,\delta_{n^a_1+\ldots+n^a_3}\right]\nn\,.
\end{align}

\medskip

Finally, the 4-point amplitude of the vector bosons is obtained as 
\begin{align}
    \mathcal{A}^{(4)}_{\text{V}}(\left\{e_i,p_i\right\})\propto \tilde{g}^4c'^3\,e^1_\mu e^2_\nu  e^3_\rho e^4_\s \,&\frac{1}{(stu)}\,s^{2c'p_{(3}\cdot\zeta_{4)}} u^{2c'p_{(2}\cdot\zeta_{3)}}t^{2c'p_{(1}\cdot\zeta_{3)}}\Big\{(tp_2-sp_3)^\mu\nn\\
    &\times (-tp_3+up_4)^\nu (-sp_2+u p_4)^\rho(-sp_2+tp_3)^\s\Big\}\, ,\label{FourDF}
\end{align}
which is manifestly crossing symmetric. As expected from computation of the 4-graviton amplitude, the above vector boson amplitude also has a different pole structure as compared to the usual Yang-Mills theory. We discuss the vector boson amplitudes and their target spacetime theory in detail in our companion paper \cite{Bagchi:2026pjx}.

\section{Conclusions}\label{sec: summary}

\subsection*{Summary}
In this work, we have focused on the intrinsic computation of the scattering amplitudes of null strings in the flipped vacuum, which realises the highest weight representation of the underlying BMS algebra on the worldsheet. Crucial to our construction was the use of the vertex operator \eqref{genvertex}, first introduced in \cite{Bagchi:2026iyu}, which involved the time-independent solution of the equations of motion $Y^\mu$ and the associated quantum number $\zeta^\mu$. Another important ingredient was the $A_0$ mode, which measures a change in the target spacetime when the $\sigma$ coordinate completes a full period. This identification naturally associates $\zeta^\mu$ with a winding number in the target spacetime. A non-trivial $A_0$ mode requires the target spacetime to have compactified directions. The inclusion of $\zeta^\mu$ in the description of null strings in the flipped vacuum thus amounts to a non-trivial change in the physical spectrum of the theory. We calculated Wightman, retarded, advanced and Feynman correlation functions between various fields. We showed that the vertex operator \eqref{genvertex} transforms like a BMS primary singlet and computed the Feynman and Wightman correlation functions between N vertex operators.

\medskip

Uncompactified null strings quantized in the flipped vacuum have a truncated spectrum, with only massless level-2 fields present \cite{Bagchi:2020fpr}. However, we have seen that when compactified directions in the target spacetime and consequently the $\zeta^\mu$ quantum number are included, the situation changes and the physical spectrum consists of massless fields that are not restricted to level-2, in keeping with observations of \cite{Banerjee:2023ekd, Banerjee:2024fbi}. Diffeomorphism and scaling invariance of the integrated vertex operator impose physical conditions at each level. Vertex operators at each level form a multiplet with a non-trivial boost matrix $\xi$, but not all components of the multiplet correspond to a physical field. The physical fields are determined by considering linear combinations of the multiplet components at each level and checking whether the combinations transform as a primary singlet. Additionally, physical fields should also satisfy the gauge condition. We find that the physical fields at level 2 are special. From the point of view of non-compact physics, it is possible to have massless physical fields at finite $\{R^a\}$. 

\medskip

Having established the necessary framework, we compute scattering amplitudes of vector bosons and symmetric tensor bosons, and find that the four-point amplitude localises on the solution of the scattering equation \eqref{eq: scattering-eqn}, which is reminiscent of the CHY formalism. Notably, the four-point amplitudes exhibit both crossing symmetry and gauge invariance. Other physical fields, such as the graviton, dilaton, vectron, and scalaron, represent specific cases of the SB field. 

\medskip

In particular, we find that the four-point graviton amplitude factorises into two three-point graviton amplitudes. In other words, only the graviton contributes to the factorisation of the tree-level graviton scattering amplitudes. We also find that the 4-point graviton amplitude has a third-order pole at $s=0$. From this analysis, we conclude that the graviton propagator depends on the momentum as $p^{-6}$. This indicates a higher-derivative effective theory of gravity in the target spacetime rather than the usual Einstein gravity. Our companion paper \cite{Bagchi:2026pjx} provides the details of the target spacetime theory. 

\medskip

A further consequence of the null string amplitude construction developed here is explored in the companion paper \cite{Bagchi:2026pjx}.  The polarization-dependent correlator (\ref{eq:EFcorrV}) of the level-one vector states is precisely the CHY one-cycle half-integrand $W_{11\ldots 1}$, while distinguished compact momentum--winding sectors generate the complementary Parke--Taylor factor.  Their pairing reproduces the pure-vector sector of the higher-derivative $(DF)^2$ gauge theory.  For factorized, transverse and traceless level-two states, the winding setup  can instead be chosen to be trivial, and the {\it same} Carrollian world-sheet directly produces the symmetric product
$W_{11\ldots 1}(\epsilon)\, W_{11\ldots 1}(\widetilde\epsilon)$,
corresponding to the six-derivative gravitational double copy of $(DF)^2$.  Thus, both kinematic copies arise on a single null-string world-sheet, without the monodromy phases familiar from tensile-string Kawai--Lewellen--Tye (KLT) relations \cite{Kawai:1985xq}.  Momentum--winding sectors with common full $D$-dimensional momenta also realize the four-point Bern--Carrasco--Johansson (BCJ) ordering relation \cite{Bern:2008qj} and, at every fixed multiplicity, the corresponding field-theory  KLT bases.  This points toward a possible Carrollian current-algebra origin of the emergent colour sector.

\subsection*{Discussions}

An interesting future direction concerns celestial \cite{Strominger:2017zoo, Pasterski:2021raf} and Carrollian approaches \cite{Bagchi:2025vri} to holography in asymptotically flat spacetimes.  In four-dimensional ambitwistor-string theories, world-sheet OPEs of conformal-primary vertex operators have been shown to reproduce the celestial OPEs of gluons and gravitons, including infinite towers of chiral descendants, while the corresponding momentum-eigenstate OPEs yield the collinear splitting functions \cite{Adamo:2021zpw}.  Since the amplitudes constructed here localize on the same scattering equations, it is natural to ask whether conformal-primary versions of the flipped-vacuum vertex operators similarly encode celestial OPE data directly on the Carrollian world-sheet.  This question is particularly suggestive in view of the representation of string amplitudes as celestial correlators and the relation between the string world-sheet and the celestial sphere in the high-energy, zero-tension regime \cite{Stieberger:2018edy,Kervyn:2025wsb}.  The latter relation is based on the saddle-point expansion of tensile-string amplitudes, whereas the flipped vacuum defines a quantum theory directly at the tensionless point.  Establishing a precise map between these descriptions could therefore provide a new bridge between null strings, high-energy string theory and celestial holography, and may reveal a world-sheet origin of celestial operator algebras and their higher-spin sectors.
A complementary perspective is provided by \cite{Stieberger:2024shv}, where string amplitudes are recast as correlators of a three-dimensional Carrollian CFT and the infinite-retarded-time limit relates the string world-sheet to the celestial sphere.  Combined with the ambitwistor-string derivation of celestial OPEs \cite{Adamo:2021zpw} and with the intrinsic CHY localization on the Carrollian world-sheet, this suggests that the flipped null string may provide a direct Carrollian world-sheet description of celestial scattering data at the exact tensionless point.

\medskip

A very natural next step is to generalise our constructions to the null superstring. The classical null superstring comes in two distinct varieties, the homogeneous \cite{Lindstrom:1990qb, Bagchi:2016yyf}, and the inhomogeneous \cite{Bagchi:2017cte}, which have different Super BMS algebras as the residual symmetry algebras on their respective null worldsheets. A careful canonical quantization along the methods of \cite{Bagchi:2020fpr} should reveal a much richer structure of vacua here, and then various consistent quantum null superstring theories can be formulated. Since vertex operators are intimately tied to the choice of vacuum, as we have seen in \cite{Bagchi:2026iyu} for the induced vacuum and here for the highest weight vacuum, the construction of vertex operators for the superstring would also be intricate. Integrated vertex operators, once constructed, would help us compute scattering amplitudes for various null superstrings. 

\bigskip \bigskip

\section*{Acknowledgements}
AB thanks Aritra Banerjee and Shankhadeep Chakrabortty for many interesting discussions and an ongoing collaboration on various aspects of tensionless strings over the last decade. 

\smallskip

AB thanks StSt for an invitation to MPI Munich and for hospitality in June 2026 during which the project was initially formulated. AB's research is partially supported by ANRF grants ANRF/ARGM/2025/000653/MTR and CRG/2022/006165. AB also gratefully acknowledges the support of the Gireesh Jankinath Chair Professorship at IIT Kanpur. 

\smallskip
SG thanks Md. Abhishek, Paolo Pergola and Biswajit Sahoo for insightful discussions. SG's research is supported by Institute postdoctoral fellowship from IIT Kanpur.

\smallskip

SRI thanks Arkachur Bhattacharjee, Emil Have, Priyadarshini Pandit, Atanu Samanta, Pushkar Soni and the participants of the ST$^4$ workshop for various discussions. SRI is grateful to the Chennai Mathematical Institute (CMI) for hospitality during the ST$^4$ workshop and to the Universit\`e Libre de Bruxelles (ULB) for hospitality during the Carrollian Physics and Geometry Workshop. SRI is supported by the Institute Assistantship at IIT Kanpur.

\smallskip

AS is supported by a FARE fellowship from IIT Kanpur.

\smallskip

This work is  supported in part by the DFG grant 508889767 {\it 
Forschungsgruppe ``Modern foundations of scattering amplitudes''}

\newpage

\section*{APPENDICES}

\appendix

\section{Correlation functions: Details}\label{App:CorrFn Details}
First, we consider the Wightman function between the fields $X^\mu$ and $X^\nu$
\begin{align}
    \left\langle X^\mu(\s_1,\t_1) X^\nu(\s_2,\t_2)\right\rangle
    &=\left\langle x^\mu x^\nu\right\rangle+\left\langle\left[\sqrt{\frac{c'}{2}}B^\mu_0\t_1+i\sqrt{\frac{c'}{2}}\sum\limits_{n>0} \frac{1}{n}\left[A^\mu_n-in\t_1B^\mu_n\right]e^{-in\s_1}\,,\right.\right.\nn\\
    &\left.\left.x^\nu+i\sqrt{\frac{c'}{2}}\sum\limits_{m<0} \frac{1}{m}\left[A^\nu_m-im\t_2 B^\nu_m\right]e^{-im\s_2}\right]\right\rangle\nn\\
    =&\,-c'\eta^{\mu\nu}\left[\frac{i\t_1e^{i\s_1}-i\t_2e^{i\s_2}}{e^{i\s_1}-e^{i\s_2}}+i\pi\,\t_{12}\,\delta\left(\s_{12}\right)\right]\,,
\end{align}
where we have used \eqref{AB annihilation} and its hermitian conjugates: $\langle0;0,0|\hat A^\mu_n=0$ and $\langle0;0,0|\hat B^\mu_n=0$ for $n<0$, in addition to \eqref{zeromode annihilation}, and set the zero-mode correlator $\left\langle x^\mu x^\nu\right\rangle=0$ \footnote{The unique value $\frac{e^{-i\s_{12}}}{1-e^{-i\s_{12}}}$ is assigned to the formally divergent series $\sum\limits_{n>0}\,{e^{-in\s_{12}}}$ through the process of analytic continuation when $\s_{12}\notin 2\pi\mathbb{Z}$.}.

\medskip

We note that this Wightman function and the corresponding Feynman correlator \eqref{XXFeynCyl} would be $\t$-translation invariant had we imposed $\langle0;0,0|\hat B^\mu_0=0$. However, in that case, the resulting correlators would not be consistent with the canonical commutation relation (assuming that the vacuum is normalizable). Hence, we conclude that the $B^\mu_0$ mode must \textbf{not} annihilate the bra-vacuum $\langle0;0,0|$.

\medskip

We shall now attempt to derive the (analytic continuations of the) Wightman functions \eqref{YX Wight}, \eqref{XY Wight}, and \eqref{YYcorr} using the mode-expansions of the fields $X^\mu$ and $Y^\mu$; this exercise will lead us to a few consistency conditions
\begin{align*}
    &\left\langle Y^\mu(\s_1,\t_1) X^\nu(\s_2,\t_2)\right\rangle\nn\\
    =&\,\left\langle y^\mu x^\nu\right\rangle-ic'\eta^{\mu\nu}\s_1-c'\eta^{\mu\nu}\sum_{n>0}\,\frac{e^{-in\s_{12}}}{-n}+i\sqrt{\frac{c'}{2}}\sum\limits_{m<0} \frac{1}{m}\left[\left\langle y^\mu A^\nu_m\right\rangle-im\t_2\left\langle y^\mu B^\nu_m\right\rangle\right]e^{-im\s_2}\,,\\
    &{}\\
    &\left\langle X^\mu(\s_1,\t_1) Y^\nu(\s_2,\t_2) \right\rangle\nn\\
    =&\,\left\langle x^\mu y^\nu\right\rangle+\sqrt{\frac{c'}{2}}\left\langle A^\mu_0 y^\nu\right\rangle\s_1+\sqrt{\frac{c'}{2}}\sum_{m\geq0}\left\langle B^\mu_m y^\nu\right\rangle\t_1e^{-im\s_1}+i\sqrt{\frac{c'}{2}}\sum_{m>0}\frac{1}{m}\left\langle A^\mu_m y^\nu\right\rangle e^{-im\s_1}\\
    &-c'\eta^{\mu\nu}\sum_{n>0}\,\frac{e^{-in\s_{12}}}{-n}\,,\\
    &{}\\
    &\left\langle Y^\mu(\s_1,\t_1) Y^\nu(\s_2,\t_2)\right\rangle\\
    =&\,\left\langle y^\mu y^\nu\right\rangle+\sqrt{\frac{c'}{2}}\left\langle B^\mu_0 y^\nu\right\rangle\s_1+i\sqrt{\frac{c'}{2}}\sum_{n<0}\frac{1}{n}\left\langle y^\mu B^\nu_n\right\rangle e^{-in\s_2}+i\sqrt{\frac{c'}{2}}\sum_{n>0}\frac{1}{n}\left\langle B^\mu_n y^\nu\right\rangle e^{-in\s_1}\,.
\end{align*}

\medskip

First, consistency with the correlator \eqref{YYcorr} requires 
\begin{align}
    \left[y^\mu\,,B^\nu_n\right]=0\,\hspace{5mm}\text{for }n\in\mathbb{Z}\,.
\end{align}
Then comparing the above expressions with the Wightman functions \eqref{YX Wight}-\eqref{XY Wight} using the cylinder to plane map \eqref{plane-cyl map}, we conclude that
\begin{align}
    \left[y^\mu\,,A^\nu_n\right]=i\sqrt{2c'}\,\eta^{\mu\nu}\delta_{n,0}\,.
\end{align}
All zero-mode correlators, such as $\left\langle y^\mu y^\nu\right\rangle\,$, $\left\langle x^\mu y^\nu\right\rangle$ are chosen as $0$ \footnote{This can be achieved by assuming that both the zero-modes $x^\mu$ and $y^\mu$ annihilate the bra-vacuum $\langle0;0,0|$.}. Finally, compatibility with the cylinder correlator \eqref{XXWightCyl} via the relation \eqref{XYrelation} demands the following integration conventions for complex logarithms
\begin{align*}
    &\int d\s_1\left[\frac{ie^{i\s_1}}{e^{i\s_1}-e^{i\s_2}}+i\pi\,\delta\left(\s_{12}\right)\right]:=\log\left(e^{i\s_1}-e^{i\s_2}\right)\,,\\
    &\int d\s_2\left[\frac{-ie^{i\s_2}}{e^{i\s_1}-e^{i\s_2}}-i\pi\,\delta\left(\s_{12}\right)\right]:=\log\left(e^{i\s_1}-e^{i\s_2}\right)\,.
\end{align*}
With these logarithm conventions, we finally have the following cylinder counterparts of the plane correlators \eqref{YX Wight}-\eqref{XY Wight}:
\begin{align}
    \left\langle Y^\mu(\s_1,\t_1) X^\nu(\s_2,\t_2)\right\rangle=\left\langle X^\mu(\s_1,\t_1) Y^\nu(\s_2,\t_2) \right\rangle=-c'\eta^{\mu\nu}\log\left(e^{i\s_1}-e^{i\s_2}\right)\,,
\end{align}
that, in particular, are not $\s$-translation invariant.

\medskip

Generalising the above derivation of the various Wightman functions and Feynman correlators using the mode-expansions \eqref{Xexp} and \eqref{Yexp}, it can be easily shown that Wick's theorem holds for both of these types of correlators on the $|0;0,0\rangle$ vacuum. E.g., we have:
\begin{align}
    &X^\mu(x_1,t_1) X^\nu(x_2,t_2)=\,:X^\mu(x_1,t_1) X^\nu(x_2,t_2):+\left\langle X^\mu(x_1,t_1) X^\nu(x_2,t_2)\right\rangle\,,\nn\\
  \Longrightarrow\hspace{2.5mm}&\hat{\mathcal{T}} X^\mu(x_1,t_1) X^\nu(x_2,t_2)=\,:X^\mu(x_1,t_1) X^\nu(x_2,t_2):+\,G^{\mu\nu}_F\left(x_1,t_1;x_2,t_2\right)\,,
\end{align}
on plane and similar relations on cylinder. Our choice that the zero-mode correlators like $\left\langle x^\mu x^\nu\right\rangle$ vanish then corresponds to setting the normal-ordered correlators to zero, e.g.:
\begin{align*}
    \left\langle\,:X^\mu(x_1,t_1) X^\nu(x_2,t_2):\right\rangle=\left\langle x^\mu x^\nu\right\rangle=0\,.
\end{align*}

\section{T-V OPEs}\label{App:OPEs}
The OPE between the scalar field $X^\mu(1)$ and the vertex operator $\mathcal V_{p,\zeta}(2)$ is given as follows 
\begin{equation}
X^\mu(1)\mathcal{V}_{p,\zeta}(2)\sim-ic'\left(p^\mu \frac{t_{12}}{x_{12}-i\epsilon t_{12}}+\zeta^\mu\log(x_{12}-i\epsilon t_{12})\right)\mathcal{V}_{p,\zeta}(2)\,. 
\end{equation}

\subsection*{Level 1}
The vertex operators at level 1 are $\lambda_\mu\dot X^\mu\mathcal V_{p,\zeta}$ and $\Xi_\mu X'^\mu\mathcal V_{p,\zeta}$. The OPEs of the vertex operators with the components of the stress tensor are given below
\begin{subequations}
    \begin{align}
        :T^t_t(1)::\lambda_{\mu}\dot X^\mu\mathcal V_{p,\zeta}(2):&\sim\lambda_\mu\left(\frac{\frac{c'p^2}{2}}{x_{12}^2}+\frac{\partial_t}{x_{12}}\right)\dot X^\mu\mathcal V_{p,\zeta}(2)\,,\label{Tttlambda}\\
        :T^t_x(1)::\lambda_{\mu}\dot X^\mu\mathcal V_{p,\zeta}(2):&\sim\lambda_{\mu}\left(\frac{1}{x_{12}^2}-\frac{c'p^2t_{12}}{x_{12}^3}+\frac{c'\zeta\cdot p}{x_{12}^2}+\frac{\partial_x}{x_{12}}-\frac{t_{12}\partial_t}{x_{12}^2}\right)\dot X^\mu\mathcal V_{p,\zeta}(2)\,.\label{Ttxlambda}
    \end{align}
\end{subequations}
To arrive at \eqref{Ttxlambda}, we have set $\lambda_\mu p^\mu=0$. The vertex operator, $\dot X^\mu\mathcal V_{p,\zeta}$ has scaling dimension $\Delta=1+c'p\cdot\zeta$.
\begin{subequations}
    \begin{align}
        :T^t_t(1)::\Xi_{\mu}X'^\mu\mathcal V_{p,\zeta}(2):&\sim\Xi_\mu\left(\frac{1}{x_{12}^2}\dot X^\mu+\frac{\frac{c'p^2}{2}}{x_{12}^2}X'^\mu+\frac{\partial_t}{x_{12}}X'^\mu\right)\mathcal V_{p,\zeta}(2)\,,\label{TttXi}\\
        :T^t_x(1)::\Xi_{\mu}X'^\mu\mathcal V_{p,\zeta}(2):&\sim\Xi_{\mu}\left(\frac{1}{x_{12}^2}X'^\mu+\frac{\partial_x}{x_{12}}X'^\mu-\frac{t_{12}\partial_t}{x_{12}^2}X'^\mu-\frac{c'p^2t_{12}}{x_{12}^3}X'^\mu\right.\nn\\&~~~~~~~~~~~~~~~~~~\left.+\frac{c'\zeta\cdot p}{x_{12}^2}X'^\mu-\frac{2t_{12}}{x_{12}^3}\dot X^\mu\right)\mathcal V_{p,\zeta}(2)\,.\label{TtxXi}
    \end{align}
\end{subequations}
To arrive at \cref{TttXi,TtxXi}, we have set $\Xi_\mu p^\mu=\Xi_\mu\zeta^\mu=0$. The vertex operator, $X'^\mu\mathcal V_{p,\zeta}$ has scaling dimension $\Delta=1+c'p\cdot\zeta$. The boost matrix $\xi$ is given as 
\begin{equation}
    \bm\xi
    \begin{pmatrix}
        \dot X^\mu\mathcal V_{p,\zeta}\\ X'^\mu\mathcal V_{p,\zeta}
    \end{pmatrix}
    =
    \begin{pmatrix}
        c'p^2/2&0\\1&c'p^2/2
    \end{pmatrix}
    \begin{pmatrix}
        \dot X^\mu\mathcal V_{p,\zeta}\\ X'^\mu\mathcal V_{p,\zeta}
    \end{pmatrix}\,.
\end{equation}

\subsection*{Level 2}
The vertex operators at level 2 are\footnote{The fifth term is set to 0 owing to the equations of motion.}
\begin{multline}
\alpha_{\mu\nu}\dot X^\mu\dot X^\nu\mathcal V_{p,\zeta}+\beta_{\mu\nu}X'^\mu X'^\nu\mathcal V_{p,\zeta}+\gamma_{\mu\nu}\dot X^\mu X'^\nu\mathcal{V}_{p,\zeta}+\tilde\gamma_{\mu\nu}X'^\mu\dot X^\nu\mathcal{V}_{p,\zeta}\\+G_{\mu\nu}\underbrace{\left(\dot X^\mu\dot X^\nu+ic'p^\mu\dot X'^\nu\right)\mathcal{V}_{p,\zeta}}_{\mathcal{V}^{\mu\nu}}+\cancelto{0}{\lambda_\mu\ddot X^\mu\mathcal{V}_{p,\zeta}}+\omega_\mu X''^\mu\mathcal{V}_{p,\zeta}+\delta_\mu\partial_t\partial_xX^\mu\mathcal{V}_{p,\zeta}\,.
\end{multline}
Here, $\alpha_{\mu\nu}$, $G_{\mu\nu}$ and $\beta_{\mu\nu}$ are symmetric. The vertex operator corresponding to $G_{\mu\nu}$ was introduced in \cite{Figueroa-OFarrill:2026igk}. The OPEs of the various vertex operators with the components of the stress tensor are given below. 
\begin{subequations}
\begin{align}
    :T^t_t(1)::\alpha_{\mu\nu}\dot X^\mu\dot X^\nu\mathcal V_{p,\zeta}(2):&\sim\alpha_{\mu\nu}\left(\frac{\frac{c'p^2}{2}}{x_{12}^2}+\frac{\partial_t}{x_{12}}\right)\dot X^\mu\dot X^\nu\mathcal V_{p,\zeta}(2)\,,\\
    :T^t_x(1)::\alpha_{\mu\nu}\dot X^\mu\dot X^\nu\mathcal V_{p,\zeta}(2):&\sim\alpha_{\mu\nu}\left(\frac{2}{x_{12}^2}-\frac{c'p^2t_{12}}{x_{12}^3}+\frac{c'\zeta\cdot p}{x_{12}^2}-\frac{t_{12}}{x_{12}^2}\partial_t\right.\nn\\&~~~~~~~~~~~~~~~~~~~~~~~~~~\left.+\frac{1}{x_{12}}\partial_x\right)\dot X^\mu\dot X^\nu\mathcal V_{p,\zeta}(2)\,.\label{Ttxalpha}
\end{align}
\end{subequations}
To arrive at \eqref{Ttxalpha}, we have set $\alpha_{\mu\nu}p^\mu=\alpha_{\mu\nu}p^\nu=0$. Therefore, the labels of $\dot X^\mu\dot X^\nu\mathcal V_{p,\zeta}$ can be read off as $\Delta=2+c'p\cdot\zeta$ and $\xi=c'p^2/2$.
\begin{subequations}
    \begin{align}
        :T^t_t(1)::\beta_{\mu\nu} X'^\mu X'^\nu\mathcal V_{p,\zeta}(2):&\sim\beta_{\mu\nu}\left(\frac{1}{x_{12}^2}(\dot X^\mu X'^\nu+X'^\mu\dot X^\nu)\right.\nn\\&~~~~~~~~~~~~~~\left.+\frac{\frac{c'p^2}{2}}{x_{12}^2}X'^\mu X'^\nu+\frac{\partial_t}{x_{12}}X'^\mu X'^\nu\right)\mathcal V_{p,\zeta}(2)\,,\label{Tttbeta}\\
    :T^t_x(1)::\beta_{\mu\nu} X'^\mu X'^\nu\mathcal V_{p,\zeta}(2):&\sim\beta_{\mu\nu}\left(\left(\frac{2}{x_{12}^2}+\frac{\partial_x}{x_{12}}-\frac{t_{12}\partial_t}{x_{12}^2}-\frac{c'p^2t_{12}}{x_{12}^3}+\frac{c'\zeta\cdot p}{x_{12}^2}\right)X'^\mu X'^\nu\right.\nn\\&~~~~~~~~~~~\left.-\frac{t_{12}}{x_{12}^3}\left(2\dot X^\mu X'^\nu+2X'^\mu\dot X^\nu\right)\right)\mathcal V_{p,\zeta}(2)\label{Ttxbeta}\,.
    \end{align}
\end{subequations}
To arrive at \cref{Tttbeta,Ttxbeta}, we have set $\eta^{\mu\nu}\beta_{\mu\nu}=p^\mu\beta_{\mu\nu}=\zeta^\mu\beta_{\mu\nu}=0$. The vertex operator, $X'^\mu X'^\nu\mathcal V_{p,\zeta}$ has scaling dimension $\Delta=2+c'p\cdot\zeta$. Clearly, $\xi$ is not a number for the level 2 operators. The exact form of the matrix will be determined after all the OPEs have been calculated. 
\begin{subequations}
    \begin{align}
        :T^t_t(1)::\gamma_{\mu\nu}\dot X^\mu X'^\nu\mathcal V_{p,\zeta}(2):&\sim\gamma_{\mu\nu}\left(\frac{\partial_t}{x_{12}}\dot X^\mu X'^\nu+\frac{1}{x_{12}^2}\dot X^\mu\dot X^\nu+\frac{\frac{c'p^2}{2}}{x_{12}^2}\dot X^\mu X'^\nu\right)\mathcal V_{p,\zeta}(2)\,,\label{Tttgamma}\\
    :T^t_x(1)::\gamma_{\mu\nu}\dot X^\mu X'^\nu\mathcal V_{p,\zeta}(2):&\sim\gamma_{\mu\nu}\left(\left(\frac{2}{x_{12}^2}-\frac{t_{12}\partial_t}{x_{12}^2}+\frac{\partial_x}{x_{12}}-\frac{c'p^2t_{12}}{x_{12}^3}+\frac{c'\zeta\cdot p}{x_{12}^2}\right)\dot X^\mu X'^\nu\right.\nn\\&~~~~~~~~~~~~~~~~~~~~~~~~~~~~~~~\left.-2\frac{t_{12}}{x_{12}^3}\dot  X^\mu\dot X^\nu\right)\mathcal V_{p,\zeta}(2)\label{Ttxgamma}\,.
    \end{align}
\end{subequations}
To arrive at \cref{Tttgamma,Ttxgamma}, we have set $\gamma_{\mu\nu}p^\nu=\eta^{\mu\nu}\gamma_{\mu\nu}=\gamma_{\mu\nu}p^\mu=\gamma_{\mu\nu}\zeta^\nu=0$. The vertex operator, $\dot X^\mu X'^\nu\mathcal V_{p,\zeta}$ has scaling dimension $\Delta=2+c'p\cdot\zeta$.
\begin{subequations}
    \begin{align}
        :T^t_t(1)::\tilde\gamma_{\mu\nu}X'^\mu \dot X^\nu\mathcal V_{p,\zeta}(2):&\sim\tilde\gamma_{\mu\nu}\left(\frac{\partial_t}{x_{12}}X'^\mu \dot X^\nu+\frac{1}{x_{12}^2}\dot X^\mu\dot X^\nu+\frac{\frac{c'p^2}{2}}{x_{12}^2}X'^\mu\dot X^\nu\right)\mathcal V_{p,\zeta}(2)\,,\label{Ttttildegamma}\\
    :T^t_x(1)::\tilde\gamma_{\mu\nu}X'^\mu \dot X^\nu\mathcal V_{p,\zeta}(2):&\sim\tilde\gamma_{\mu\nu}\left(\left(\frac{2}{x_{12}^2}-\frac{t_{12}\partial_t}{x_{12}^2}+\frac{\partial_x}{x_{12}}-\frac{c'p^2t_{12}}{x_{12}^3}+\frac{c'\zeta\cdot p}{x_{12}^2}\right)X'^\mu \dot X^\nu\right.\nn\\&~~~~~~~~~~~~~~~~~~~~~~~~~~~~~~~~~~~~~~\left.-2\frac{t_{12}}{x_{12}^3}\dot  X^\mu\dot X^\nu\right)\mathcal V_{p,\zeta}(2)\label{Ttxtildegamma}\,.
    \end{align}
\end{subequations}
To arrive at \cref{Ttttildegamma,Ttxtildegamma}, we have set $\tilde\gamma_{\mu\nu}p^\mu=\eta^{\mu\nu}\tilde\gamma_{\mu\nu}=\tilde\gamma_{\mu\nu}p^\nu=\tilde\gamma_{\mu\nu}\zeta^\mu=0$. The vertex operator, $X'^\mu \dot X^\nu\mathcal V_{p,\zeta}$ has scaling dimension $\Delta=2+c'p\cdot\zeta$.

\begin{subequations}
    \begin{align}
    :T^t_t(1)::G_{\mu\nu}\mathcal{V}^{\mu\nu}(2):&\sim G_{\mu\nu}\left(\frac{\partial_t}{x_{12}}+\frac{\frac{c'p^2}{2}}{x_{12}^2}\right)\mathcal V^{\mu\nu}(2)\,,\label{eq: TttG}\\
    :T^t_x(1)::G_{\mu\nu}\mathcal V^{\mu\nu}(2):&\sim G_{\mu\nu}\left(\frac{2}{x_{12}^2}+\frac{c'p\cdot\zeta}{x_{12}^2}-\frac{c'p^2t_{12}}{x_{12}^3}-\frac{t_{12}\partial_t}{x_{12}^2}+\frac{\partial_x}{x_{12}}\right)\mathcal V^{\mu\nu}(2)\label{eq: TtxG}\,.
    \end{align}
\end{subequations}
To arrive at \eqref{eq: TtxG}, we have set $G_{\mu\nu}p^\mu p^\nu=0$, which exactly matches with the condition given in \cite{Figueroa-OFarrill:2026igk}.
\begin{subequations}
    \begin{align}
        :T^t_t(1)::\omega_\mu X''^\mu\mathcal V_{p,\zeta}(2):&\sim\omega_\mu\left(-\frac{2ic'}{x_{12}^4}p^\mu+\frac{2}{x_{12}^3}\dot X^\mu+\frac{2}{x_{12}^2}\dot X'^\mu+\frac{\partial_t}{x_{12}}X''^\mu+\frac{\frac{c'p^2}{2}}{x_{12}^2}X''^\mu\right)\mathcal V_{p,\zeta}(2)\,,\label{Tttomega}\\
        :T^t_x(1)::\omega_\mu X''^\mu\mathcal V_{p,\zeta}(2):&\sim\omega_\mu\left(\frac{8ic't_{12}}{x_{12}^5}p^\mu-\frac{2ic'}{x_{12}^4}\zeta^\mu+\frac{2}{x_{12}^2}X''^\mu+\frac{\partial_x}{x_{12}}X''^\mu-\frac{c'p^2t_{12}}{x_{12}^3}X''^\mu\right.\nn\\&\left.+\frac{c'\zeta\cdot p}{x_{12}^2}X''^\mu-\frac{4t_{12}}{x_{12}^3}\dot X'^\mu-\frac{t_{12}\partial_t}{x_{12}^2}X''^\mu-\frac{6t_{12}}{x_{12}^4}\dot X^\mu+\frac{2}{x_{12}^3}X'^\mu\right)\mathcal{V}_{p,\zeta}(2)\,.\label{Ttxomega}
    \end{align}
\end{subequations}
Clearly, the operator $X''^\mu\mathcal V_{p,\zeta}$ does not transform as a primary operator unless $\omega_\mu p^\mu=\omega_\mu\dot X^\mu=\zeta^\mu\omega_\mu=X'^\mu\omega_\mu=0\,.$ These conditions force $\omega_\mu=0$.
\begin{subequations}
    \begin{align}
        :T^t_t(1)::\delta_\mu \dot X'^\mu\mathcal V_{p,\zeta}(2):&\sim\delta_\mu\left(\frac{\partial_t}{x_{12}}\dot X'^\mu+\frac{\frac{c'p^2}{2}}{x_{12}^2}\dot X'^\mu\right)\mathcal V_{p,\zeta}(2)\,,\label{Tttdelta}\\
        :T^t_x(1)::\delta_\mu \dot X'^\mu\mathcal V_{p,\zeta}(2):&\sim\delta_\mu\left(-\frac{2ic'}{x_{12}^4}p^\mu+\frac{2}{x_{12}^3}\dot X^\mu+\frac{2}{x_{12}^2}\dot X'^\mu+\frac{\partial_x}{x_{12}}\dot X'^\mu-\frac{t_{12}\partial_t}{x_{12}^2}\dot X'^\mu\right.\nn\\&~~~~~~~~~~~~~~~~~~\left.-\frac{c'p^2t_{12}}{x_{12}^3}\dot X'^\mu+\frac{c'\zeta\cdot p}{x_{12}^2}\dot X'^\mu\right)\mathcal{V}_{p,\zeta}(2)\,.\label{Ttxdelta}
    \end{align}
\end{subequations}
Clearly, the operator $\dot X'^\mu\mathcal V_{p,\zeta}$ does not transform as a primary operator unless $\delta_\mu p^\mu=\delta_\mu\dot X^\mu=0\,.$ These conditions force $\delta_\mu=0$.

\medskip

Now, we can sketch out the $\xi-$matrix for the first four vertex operators discussed in this section. In multiplet form, this is given as 
\begin{equation}
   \bm\xi 
    \begin{pmatrix}
        \dot X^\mu\dot X^\nu\mathcal V_{p,\zeta}\\\left(\dot X^\mu\,,X'^\nu\right)_+\mathcal V_{p,\zeta}\\X'^\mu X'^\nu\mathcal V_{p,\zeta}
    \end{pmatrix}
    =
    \begin{pmatrix}
    c'p^2/2&0&0\\1&c'p^2/2&0\\0&1&c'p^2/2        
    \end{pmatrix}\begin{pmatrix}
        \dot X^\mu\dot X^\nu\mathcal V_{p,\zeta}\\\left(\dot X^\mu\,,X'^\nu\right)_+\mathcal V_{p,\zeta}\\X'^\mu X'^\nu\mathcal V_{p,\zeta}
    \end{pmatrix}
\end{equation}
and $\bm\xi\left(\dot X^\mu\,,X'^\nu\right)_-\mathcal V_{p,\zeta}=\frac{c'p^2}{2}\left(\dot X^\mu\,,X'^\nu\right)_-\mathcal V_{p,\zeta}$.

\section{\texorpdfstring{$B-$ Field analysis}{B-Field Analysis}}\label{sec: B Field analysis}
The vertex operator for the $B-$field is given as $\mathcal{V}^B=f_{\mu\nu}\partial_tX^\mu\partial_xX^\nu\mathcal V_{p,\zeta}$, where $f_{\mu\nu}$ is an antisymmetric tensor and satisfies $f_{\mu\nu}p^\mu=f_{\mu\nu}\zeta^\mu=0$. Since we are dealing with level-2 states, we can directly work with $\zeta^\mu=0$ (see \cref{sec: The physical fields}). The two-point correlator of the $B-$field, unlike the physical fields in level-1 and other physical fields in level-2, does not vanish on-shell. The two-point correlator of the $B-$field can be calculated as follows
\begin{equation}
    \braket{\mathcal V^B(1)\mathcal V^B(2)}=\frac{c'^2}{x_{12}^4}f_{\mu\nu}^1f_{\rho\sigma}^2\eta^{\mu\sigma}\eta^{\nu\rho}\delta^{(D)}(p_1+p_2)\,.
\end{equation}
\subsection*{Graviton-Graviton-B Field Scattering}
The correlation function between two gravitons and a B field is
\begin{multline}
    \braket{\mathcal V^G(1)\mathcal V^G(2)\mathcal V^B(3)}=-2c'e^1_{\mu\nu}e^2_{\rho\sigma}f_{\alpha\beta}^3
    f_1^\mu f_2^\rho f_3^\alpha\left(\frac{\eta^{\nu\beta}f_2^\sigma}{x_{13}^2}+\frac{\eta^{\sigma\beta}f_1^\nu}{x_{23}^2}\right)\\
    \exp\left(ic'\sum_{i<j}^{j=3}p_i\cdot p_j\frac{t^E_{ij}}{x_{ij}}\right)\delta^{(D)}\left(\sum_{i=1}^3p_i\right)\,.
\end{multline}
The corresponding amplitude in the gauge $x_1=a$, $x_2=0$, $x_3=1$, $t_i=0$ and $a\to\infty$ is given as 
\begin{equation}\label{eq: GGB amplitude}
\mathcal A_{GGB}=2c'^5e^1_{\mu\nu}\,p_3^\mu\, e^2_{\rho\sigma}\,p_1^\rho\,f^3_{\alpha\beta}\,p_2^\alpha\left(p_3^\sigma\eta^{\nu\beta}-p_3^\nu\eta^{\sigma\beta}\right)\delta^{(D)}\left(\sum_{i=1}^3p_i\right)\,.
\end{equation}
To obtain the above-mentioned amplitude, we have multiplied by the conformal factor $a^4$.

\newpage
\bibliographystyle{JHEP}
\bibliography{biblio.bib}

@article{Gomis:2000bd,
    author = "Gomis, Jaume and Ooguri, Hirosi",
    title = "{Nonrelativistic closed string theory}",
    eprint = "hep-th/0009181",
    archivePrefix = "arXiv",
    reportNumber = "CALT-68-2298, CITUSC-00-055",
    doi = "10.1063/1.1372697",
    journal = "J. Math. Phys.",
    volume = "42",
    pages = "3127--3151",
    year = "2001"
}

@article{Sheikh-Jabbari:2026tpf,
    author = "Sheikh-Jabbari, M. M. and Yavartanoo, H.",
    title = "{Null strings gauged and reloaded, II: Consistent classical treatment of the null strings}",
    eprint = "2605.26822",
    archivePrefix = "arXiv",
    primaryClass = "hep-th",
    doi = "10.1016/j.physletb.2026.140775",
    journal = "Phys. Lett. B",
    volume = "880",
    pages = "140775",
    year = "2026"
}

@article{Duary:2026lmk,
    author = "Duary, Sarthak and Maji, Sourav",
    title = "{BRST quantization of Carroll-Weyl gauged null strings}",
    eprint = "2608.02731",
    archivePrefix = "arXiv",
    primaryClass = "hep-th",
    month = "8",
    year = "2026"
}

@article{Chen:2026cau,
    author = "Chen, Bin and Hu, Zezhou",
    title = "{Quantum Anomalies of Tensionless Bosonic Strings}",
    eprint = "2608.02987",
    archivePrefix = "arXiv",
    primaryClass = "hep-th",
    month = "8",
    year = "2026"
}

@article{Chen:2023esw,
    author = "Chen, Bin and Hu, Zezhou and Yu, Zhe-fei and Zheng, Yu-fan",
    title = "{Path-integral quantization of tensionless (super) string}",
    eprint = "2302.05975",
    archivePrefix = "arXiv",
    primaryClass = "hep-th",
    doi = "10.1007/JHEP08(2023)133",
    journal = "JHEP",
    volume = "08",
    pages = "133",
    year = "2023"
}

@article{Bagchi:2026pjx,
    author = "Bagchi, Arjun and Grover, Sachin and Rajesh Iyer, Sharang and Saha, Amartya and Stieberger, Stephan",
    title = "{Double Copy from the Flipped Null String}",
    eprint = "2609.07846",
    archivePrefix = "arXiv",
    primaryClass = "hep-th",
    month = "9",
    year = "2026"
}

@article{Figueroa-OFarrill:2025njv,
    author = "Figueroa-O'Farrill, Jos{\'e} and Have, Emil and Obers, Niels A.",
    title = "{Quantum carrollian bosonic strings}",
    eprint = "2509.04397",
    archivePrefix = "arXiv",
    primaryClass = "hep-th",
    reportNumber = "NORDITA 2025-048",
    doi = "10.1007/JHEP07(2026)098",
    journal = "JHEP",
    volume = "07",
    pages = "098",
    year = "2026"
}

@article{Duary:2025hdb,
    author = "Duary, Sarthak and Maji, Sourav",
    title = "{From closed to open strings: the tensionless route in Kalb-Ramond background and noncommutativity}",
    eprint = "2511.20917",
    archivePrefix = "arXiv",
    primaryClass = "hep-th",
    month = "11",
    year = "2025"
}

@article{Bern:2008qj,
    author = "Bern, Z. and Carrasco, J. J. M. and Johansson, Henrik",
    title = "{New Relations for Gauge-Theory Amplitudes}",
    eprint = "0805.3993",
    archivePrefix = "arXiv",
    primaryClass = "hep-ph",
    reportNumber = "UCLA-07-TEP-15",
    doi = "10.1103/PhysRevD.78.085011",
    journal = "Phys. Rev. D",
    volume = "78",
    pages = "085011",
    year = "2008"
}

@article{Stieberger:2024shv,
    author = "Stieberger, Stephan and Taylor, Tomasz R. and Zhu, Bin",
    title = "{Carrollian Amplitudes from Strings}",
    eprint = "2402.14062",
    archivePrefix = "arXiv",
    primaryClass = "hep-th",
    doi = "10.1007/JHEP04(2024)127",
    journal = "JHEP",
    volume = "04",
    pages = "127",
    year = "2024"
}

@article{Lee:2017utr,
    author = "Lee, Kanghoon and Rey, Soo-Jong and Rosabal, J. A.",
    title = "{A string theory which isn{\textquoteright}t about strings}",
    eprint = "1708.05707",
    archivePrefix = "arXiv",
    primaryClass = "hep-th",
    doi = "10.1007/JHEP11(2017)172",
    journal = "JHEP",
    volume = "11",
    pages = "172",
    year = "2017"
}

@article{Casali:2017mss,
    author = "Casali, Eduardo and Tourkine, Piotr",
    title = "{Windings of twisted strings}",
    eprint = "1710.01241",
    archivePrefix = "arXiv",
    primaryClass = "hep-th",
    doi = "10.1103/PhysRevD.97.061902",
    journal = "Phys. Rev. D",
    volume = "97",
    number = "6",
    pages = "061902",
    year = "2018"
}

@article{Adamo:2021zpw,
    author = "Adamo, Tim and Bu, Wei and Casali, Eduardo and Sharma, Atul",
    title = "{Celestial operator products from the worldsheet}",
    eprint = "2111.02279",
    archivePrefix = "arXiv",
    primaryClass = "hep-th",
    doi = "10.1007/JHEP06(2022)052",
    journal = "JHEP",
    volume = "06",
    pages = "052",
    year = "2022"
}

@article{Sheikh-Jabbari:2026vqh,
    author = "Sheikh-Jabbari, M. M. and Yavartanoo, H.",
    title = "{Null strings gauged and reloaded, I: Null strings have Carroll-Weyl gauge symmetry}",
    eprint = "2605.25817",
    archivePrefix = "arXiv",
    primaryClass = "hep-th",
    doi = "10.1016/j.physletb.2026.140824",
    journal = "Phys. Lett. B",
    volume = "880",
    pages = "140824",
    year = "2026"
}

@article{Saha:2022gjw,
    author = "Saha, Amartya",
    title = "{Intrinsic approach to 1 + 1D Carrollian Conformal Field Theory}",
    eprint = "2207.11684",
    archivePrefix = "arXiv",
    primaryClass = "hep-th",
    doi = "10.1007/JHEP12(2022)133",
    journal = "JHEP",
    volume = "12",
    pages = "133",
    year = "2022"
}

@article{Gross:1987ar,
    author = "Gross, David J. and Mende, Paul F.",
    title = "{String Theory Beyond the Planck Scale}",
    reportNumber = "PUPT-1067",
    doi = "10.1016/0550-3213(88)90390-2",
    journal = "Nucl. Phys. B",
    volume = "303",
    pages = "407--454",
    year = "1988"
}

@article{Figueroa-OFarrill:2026igk,
    author = "Figueroa-O'Farrill, Jos{\'e} M. and Vishwa, Girish S.",
    title = "{The spectrum of the bosonic ambitwistor string revisited}",
    eprint = "2606.05500",
    archivePrefix = "arXiv",
    primaryClass = "hep-th",
    month = "6",
    year = "2026"
}

@article{Cachazo:2013gna,
    author = "Cachazo, Freddy and He, Song and Yuan, Ellis Ye",
    title = "{Scattering equations and Kawai-Lewellen-Tye orthogonality}",
    eprint = "1306.6575",
    archivePrefix = "arXiv",
    primaryClass = "hep-th",
    doi = "10.1103/PhysRevD.90.065001",
    journal = "Phys. Rev. D",
    volume = "90",
    number = "6",
    pages = "065001",
    year = "2014"
}

@article{Bagchi:2026iyu,
    author = "Bagchi, Arjun and Grover, Sachin and Rajesh Iyer, Sharang and Saha, Amartya",
    title = "{High energy scattering and null strings}",
    eprint = "2603.26910",
    archivePrefix = "arXiv",
    primaryClass = "hep-th",
    month = "3",
    year = "2026"
}

@article{Kervyn:2025wsb,
    author = "Kervyn, Xavier and Stieberger, Stephan",
    title = "{High energy string theory and the celestial sphere}",
    eprint = "2504.13738",
    archivePrefix = "arXiv",
    primaryClass = "hep-th",
    doi = "10.1007/JHEP09(2025)044",
    journal = "JHEP",
    volume = "09",
    pages = "044",
    year = "2025"
}

@book{Strominger:2017zoo,
    author = "Strominger, Andrew",
    title = "{Lectures on the Infrared Structure of Gravity and Gauge Theory}",
    eprint = "1703.05448",
    archivePrefix = "arXiv",
    primaryClass = "hep-th",
    isbn = "978-0-691-17973-5",
    publisher = "Princeton University Press",
    year = "2018"
}

@inproceedings{Pasterski:2021raf,
    author = "Pasterski, Sabrina and Pate, Monica and Raclariu, Ana-Maria",
    title = "{Celestial Holography}",
    booktitle = "{Snowmass 2021}",
    eprint = "2111.11392",
    archivePrefix = "arXiv",
    primaryClass = "hep-th",
    month = "11",
    year = "2021"
}

@article{Bagchi:2012yk,
      author         = "Bagchi, Arjun and Detournay, Stephane and Grumiller,
                        Daniel",
      title          = "{Flat-Space Chiral Gravity}",
      journal        = "Phys. Rev. Lett.",
      volume         = "109",
      year           = "2012",
      pages          = "151301",
      doi            = "10.1103/PhysRevLett.109.151301",
      eprint         = "1208.1658",
      archivePrefix  = "arXiv",
      primaryClass   = "hep-th",
      SLACcitation   = "%%CITATION = ARXIV:1208.1658;%%"
}

@article{Bagchi:2019cay,
    author = "Bagchi, Arjun and Banerjee, Aritra and Parekh, Pulastya",
    title = "{Tensionless Path from Closed to Open Strings}",
    eprint = "1905.11732",
    archivePrefix = "arXiv",
    primaryClass = "hep-th",
    doi = "10.1103/PhysRevLett.123.111601",
    journal = "Phys. Rev. Lett.",
    volume = "123",
    number = "11",
    pages = "111601",
    year = "2019"
}

@article{Barnich:2014kra,
      author         = "Barnich, Glenn and Oblak, Blagoje",
      title          = "{Notes on the BMS group in three dimensions: I. Induced
                        representations}",
      journal        = "JHEP",
      volume         = "06",
      year           = "2014",
      pages          = "129",
      doi            = "10.1007/JHEP06(2014)129",
      eprint         = "1403.5803",
      archivePrefix  = "arXiv",
      primaryClass   = "hep-th",
      SLACcitation   = "%%CITATION = ARXIV:1403.5803;%%"
}

@article{Campoleoni:2016vsh,
      author         = "Campoleoni, Andrea and Gonzalez, Hernan A. and Oblak,
                        Blagoje and Riegler, Max",
      title          = "{BMS Modules in Three Dimensions}",
      booktitle      = "{Proceedings, International Workshop on Higher Spin Gauge
                        Theories: Singapore, Singapore, November 4-6, 2015}",
      journal        = "Int. J. Mod. Phys.",
      volume         = "A31",
      year           = "2016",
      number         = "12",
      pages          = "1650068",
      doi            = "10.1142/S0217751X16500688, 10.1142/9789813144101_0011",
      eprint         = "1603.03812",
      archivePrefix  = "arXiv",
      primaryClass   = "hep-th",
      SLACcitation   = "%%CITATION = ARXIV:1603.03812;%%"
}

@article{Bagchi:2009pe,
	Archiveprefix = {arXiv},
	Author = {Bagchi, Arjun and Gopakumar, Rajesh and Mandal, Ipsita and Miwa, Akitsugu},
	Doi = {10.1007/JHEP08(2010)004},
	Eprint = {0912.1090},
	Journal = {JHEP},
	Pages = {004},
	Primaryclass = {hep-th},
	Reportnumber = {HRI-ST-0923},
	Slaccitation = {%%CITATION = ARXIV:0912.1090;%%},
	Title = {{GCA in 2d}},
	Volume = {08},
	Year = {2010}}

@article{Bagchi:2009ca,
	Archiveprefix = {arXiv},
	Author = {Bagchi, Arjun and Mandal, Ipsita},
	Doi = {10.1016/j.physletb.2009.04.030},
	Eprint = {0903.4524},
	Journal = {Phys. Lett.},
	Pages = {393-397},
	Primaryclass = {hep-th},
	Reportnumber = {HRI-ST-0910},
	Slaccitation = {%%CITATION = ARXIV:0903.4524;%%},
	Title = {{On Representations and Correlation Functions of Galilean Conformal Algebras}},
	Volume = {B675},
	Year = {2009}}

@article{Barnich:2006av,
	Archiveprefix = {arXiv},
	Author = {Barnich, Glenn and Compere, Geoffrey},
	Doi = {10.1088/0264-9381/24/5/F01, 10.1088/0264-9381/24/11/C01},
	Eprint = {gr-qc/0610130},
	Journal = {Class. Quant. Grav.},
	Pages = {F15-F23},
	Primaryclass = {gr-qc},
	Reportnumber = {ULB-TH-06-08},
	Slaccitation = {%%CITATION = GR-QC/0610130;%%},
	Title = {{Classical central extension for asymptotic symmetries at null infinity in three spacetime dimensions}},
	Volume = {24},
	Year = {2007}}

@article{Bagchi:2013bga,
	Archiveprefix = {arXiv},
	Author = {Bagchi, Arjun},
	Doi = {10.1007/JHEP05(2013)141},
	Eprint = {1303.0291},
	Journal = {JHEP},
	Pages = {141},
	Primaryclass = {hep-th},
	Reportnumber = {MIT-CTP-4445, EMPG-13-02},
	Slaccitation = {%%CITATION = ARXIV:1303.0291;%%},
	Title = {{Tensionless Strings and Galilean Conformal Algebra}},
	Volume = {05},
	Year = {2013}}

@article{Bagchi:2015nca,
	Archiveprefix = {arXiv},
	Author = {Bagchi, Arjun and Chakrabortty, Shankhadeep and Parekh, Pulastya},
	Doi = {10.1007/JHEP01(2016)158},
	Eprint = {1507.04361},
	Journal = {JHEP},
	Pages = {158},
	Primaryclass = {hep-th},
	Reportnumber = {MIT-CTP-4690},
	Slaccitation = {%%CITATION = ARXIV:1507.04361;%%},
	Title = {{Tensionless Strings from Worldsheet Symmetries}},
	Volume = {01},
	Year = {2016}}

@article{Bagchi:2016yyf,
	Archiveprefix = {arXiv},
	Author = {Bagchi, Arjun and Chakrabortty, Shankhadeep and Parekh, Pulastya},
	Doi = {10.1007/JHEP10(2016)113},
	Eprint = {1606.09628},
	Journal = {JHEP},
	Pages = {113},
	Primaryclass = {hep-th},
	Reportnumber = {MIT-CTP-4816},
	Slaccitation = {%%CITATION = ARXIV:1606.09628;%%},
	Title = {{Tensionless Superstrings: View from the Worldsheet}},
	Volume = {10},
	Year = {2016}}

@article{Mason:2013sva,
    author = "Mason, Lionel and Skinner, David",
    title = "{Ambitwistor strings and the scattering equations}",
    eprint = "1311.2564",
    archivePrefix = "arXiv",
    primaryClass = "hep-th",
    doi = "10.1007/JHEP07(2014)048",
    journal = "JHEP",
    volume = "07",
    pages = "048",
    year = "2014"
}

@article{Lindstrom:1990qb,
    author = "Lindstrom, U. and Sundborg, B. and Theodoridis, G.",
    title = "{The Zero tension limit of the superstring}",
    reportNumber = "USITP-90-12",
    doi = "10.1016/0370-2693(91)91726-C",
    journal = "Phys. Lett. B",
    volume = "253",
    pages = "319--323",
    year = "1991"
}

@article{Schild:1976vq,
      author         = "Schild, Alfred",
      title          = "{Classical Null Strings}",
      journal        = "Phys. Rev.",
      volume         = "D16",
      year           = "1977",
      pages          = "1722",
      doi            = "10.1103/PhysRevD.16.1722",
      reportNumber   = "PRINT-76-0491 (TEXAS), ANL-HEP-PR-77-23",
      SLACcitation   = "%%CITATION = PHRVA,D16,1722;%%"
}

@article{Isberg:1993av,
      author         = "Isberg, J. and Lindstrom, U. and Sundborg, B. and
                        Theodoridis, G.",
      title          = "{Classical and quantized tensionless strings}",
      journal        = "Nucl. Phys.",
      volume         = "B411",
      year           = "1994",
      pages          = "122-156",
      doi            = "10.1016/0550-3213(94)90056-6",
      eprint         = "hep-th/9307108",
      archivePrefix  = "arXiv",
      primaryClass   = "hep-th",
      reportNumber   = "USITP-93-12",
      SLACcitation   = "%%CITATION = HEP-TH/9307108;%%"
}

@article{Bagchi:2017cte,
      author         = "Bagchi, Arjun and Banerjee, Aritra and Chakrabortty,
                        Shankhadeep and Parekh, Pulastya",
      title          = "{Inhomogeneous Tensionless Superstrings}",
      journal        = "JHEP",
      volume         = "02",
      year           = "2018",
      pages          = "065",
      doi            = "10.1007/JHEP02(2018)065",
      eprint         = "1710.03482",
      archivePrefix  = "arXiv",
      primaryClass   = "hep-th",
      SLACcitation   = "%%CITATION = ARXIV:1710.03482;%%"
}

@article{Banerjee:2023ekd,
    author = "Banerjee, Aritra and Chatterjee, Ritankar and Pandit, Priyadarshini",
    title = "{Tensionless tales of compactification}",
    eprint = "2307.01275",
    archivePrefix = "arXiv",
    primaryClass = "hep-th",
    doi = "10.1007/JHEP09(2023)050",
    journal = "JHEP",
    volume = "09",
    pages = "050",
    year = "2023"
}

@article{Banerjee:2024fbi,
    author = "Banerjee, Aritra and Chatterjee, Ritankar and Pandit, Priyadarshini",
    title = "{Tensionless strings in a Kalb-Ramond background}",
    eprint = "2404.01385",
    archivePrefix = "arXiv",
    primaryClass = "hep-th",
    doi = "10.1007/JHEP06(2024)067",
    journal = "JHEP",
    volume = "06",
    pages = "067",
    year = "2024"
}

@article{Bagchi:2020ats,
    author = "Bagchi, Arjun and Banerjee, Aritra and Chakrabortty, Shankhadeep",
    title = "{Rindler Physics on the String Worldsheet}",
    eprint = "2009.01408",
    archivePrefix = "arXiv",
    primaryClass = "hep-th",
    doi = "10.1103/PhysRevLett.126.031601",
    journal = "Phys. Rev. Lett.",
    volume = "126",
    number = "3",
    pages = "031601",
    year = "2021"
}

@article{Bagchi:2020fpr,
    author = "Bagchi, Arjun and Banerjee, Aritra and Chakrabortty, Shankhadeep and Dutta, Sudipta and Parekh, Pulastya",
    title = "{A tale of three \textemdash{} tensionless strings and vacuum structure}",
    eprint = "2001.00354",
    archivePrefix = "arXiv",
    primaryClass = "hep-th",
    doi = "10.1007/JHEP04(2020)061",
    journal = "JHEP",
    volume = "04",
    pages = "061",
    year = "2020"
}

@article{Gross:1987kza,
    author = "Gross, David J. and Mende, Paul F.",
    title = "{The High-Energy Behavior of String Scattering Amplitudes}",
    reportNumber = "PUPT-1062",
    doi = "10.1016/0370-2693(87)90355-8",
    journal = "Phys. Lett. B",
    volume = "197",
    pages = "129--134",
    year = "1987"
}

@article{Bagchi:2021rfw,
    author = "Bagchi, Arjun and Mandlik, Mangesh and Sharma, Punit",
    title = "{Tensionless tales: vacua and critical dimensions}",
    eprint = "2105.09682",
    archivePrefix = "arXiv",
    primaryClass = "hep-th",
    doi = "10.1007/JHEP08(2021)054",
    journal = "JHEP",
    volume = "08",
    pages = "054",
    year = "2021"
}

@article{Casali:2016atr,
    author = "Casali, Eduardo and Tourkine, Piotr",
    title = "{On the null origin of the ambitwistor string}",
    eprint = "1606.05636",
    archivePrefix = "arXiv",
    primaryClass = "hep-th",
    doi = "10.1007/JHEP11(2016)036",
    journal = "JHEP",
    volume = "11",
    pages = "036",
    year = "2016"
}

@article{Casali:2017zkz,
    author = "Casali, Eduardo and Herfray, Yannick and Tourkine, Piotr",
    title = "{The complex null string, Galilean conformal algebra and scattering equations}",
    eprint = "1707.09900",
    archivePrefix = "arXiv",
    primaryClass = "hep-th",
    reportNumber = "DAMTP-2017-32",
    doi = "10.1007/JHEP10(2017)164",
    journal = "JHEP",
    volume = "10",
    pages = "164",
    year = "2017"
}

@article{Bagchi:2018wsn,
    author = "Bagchi, Arjun and Banerjee, Aritra and Chakrabortty, Shankhadeep and Parekh, Pulastya",
    title = "{Exotic Origins of Tensionless Superstrings}",
    eprint = "1811.10877",
    archivePrefix = "arXiv",
    primaryClass = "hep-th",
    doi = "10.1016/j.physletb.2019.135139",
    journal = "Phys. Lett. B",
    volume = "801",
    pages = "135139",
    year = "2020"
}

@article{Lindstrom:2026zno,
    author = {Lindstr{\"o}m, Ulf},
    title = "{The conformal null string in $d+2$ and $d$ dimensions}",
    eprint = "2606.22498",
    archivePrefix = "arXiv",
    primaryClass = "hep-th",
    reportNumber = "Uppsala University, Theoretical Physics UUITP-15/26",
    month = "6",
    year = "2026"
}

@article{Hao:2021urq,
    author = "Hao, Peng-xiang and Song, Wei and Xie, Xianjin and Zhong, Yuan",
    title = "{BMS-invariant free scalar model}",
    eprint = "2111.04701",
    archivePrefix = "arXiv",
    primaryClass = "hep-th",
    doi = "10.1103/PhysRevD.105.125005",
    journal = "Phys. Rev. D",
    volume = "105",
    number = "12",
    pages = "125005",
    year = "2022"
}

@article{Bagchi:2024qsb,
    author = "Bagchi, Arjun and Chakraborty, Pronoy and Chakrabortty, Shankhadeep and Fredenhagen, Stefan and Grumiller, Daniel and Pandit, Priyadarshini",
    title = "{Boundary Carrollian Conformal Field Theories and Open Null Strings}",
    eprint = "2409.01094",
    archivePrefix = "arXiv",
    primaryClass = "hep-th",
    reportNumber = "TUW-24-05",
    doi = "10.1103/PhysRevLett.134.071604",
    journal = "Phys. Rev. Lett.",
    volume = "134",
    number = "7",
    pages = "071604",
    year = "2025"
}

@article{Stieberger:2018edy,
    author = "Stieberger, Stephan and Taylor, Tomasz R.",
    title = "{Strings on Celestial Sphere}",
    eprint = "1806.05688",
    archivePrefix = "arXiv",
    primaryClass = "hep-th",
    reportNumber = "MPP-2018-136",
    doi = "10.1016/j.nuclphysb.2018.08.019",
    journal = "Nucl. Phys. B",
    volume = "935",
    pages = "388--411",
    year = "2018"
}

@article{Kawai:1985xq,
    author = "Kawai, H. and Lewellen, D. C. and Tye, S. H. H.",
    title = "{A Relation Between Tree Amplitudes of Closed and Open Strings}",
    reportNumber = "CLNS-85/667",
    doi = "10.1016/0550-3213(86)90362-7",
    journal = "Nucl. Phys. B",
    volume = "269",
    pages = "1--23",
    year = "1986"
}

@article{Bagchi:2025vri,
    author = "Bagchi, Arjun and Banerjee, Aritra and Dhivakar, Prateksh and Mondal, Saikat and Shukla, Ashish",
    title = "{The Carrollian Kaleidoscope}",
    eprint = "2506.16164",
    archivePrefix = "arXiv",
    primaryClass = "hep-th",
    month = "6",
    year = "2025"
}

@article{Bagchi:2025jgu,
    author = "Bagchi, Arjun and Chakrabortty, Shankhadeep and Chakraborty, Pronoy and Chatterjee, Ritankar and Pandit, Priyadarshini",
    title = "{Boundary Carroll CFTs: SUSY and superstrings}",
    eprint = "2508.20165",
    archivePrefix = "arXiv",
    primaryClass = "hep-th",
    doi = "10.1007/JHEP12(2025)146",
    journal = "JHEP",
    volume = "12",
    pages = "146",
    year = "2025"
}

@article{Bagchi:2026wcu,
    author = "Bagchi, Arjun and Banerjee, Aritra and Chatterjee, Ritankar and Pandit, Priyadarshini",
    title = "{The tensionless lives of null strings}",
    eprint = "2601.20959",
    archivePrefix = "arXiv",
    primaryClass = "hep-th",
    doi = "10.1016/j.physrep.2026.05.001",
    journal = "Phys. Rept.",
    volume = "1185",
    pages = "1--91",
    year = "2026"
}

@article{Cachazo:2013hca,
    author = "Cachazo, Freddy and He, Song and Yuan, Ellis Ye",
    title = "{Scattering of Massless Particles in Arbitrary Dimensions}",
    eprint = "1307.2199",
    archivePrefix = "arXiv",
    primaryClass = "hep-th",
    doi = "10.1103/PhysRevLett.113.171601",
    journal = "Phys. Rev. Lett.",
    volume = "113",
    number = "17",
    pages = "171601",
    year = "2014"
}

@article{LipinskiJusinskas:2019cej,
    author = "Lipinski Jusinskas, Renann",
    title = "{Chiral strings, the sectorized description and their integrated vertex operators}",
    eprint = "1909.04069",
    archivePrefix = "arXiv",
    primaryClass = "hep-th",
    doi = "10.1007/JHEP12(2019)143",
    journal = "JHEP",
    volume = "12",
    pages = "143",
    year = "2019"
}

@article{Siegel:2015axg,
    author = "Siegel, W.",
    title = "{Amplitudes for left-handed strings}",
    eprint = "1512.02569",
    archivePrefix = "arXiv",
    primaryClass = "hep-th",
    reportNumber = "YITP-SB-15-45",
    month = "12",
    year = "2015"
}
\end{document}